\documentclass[reprint,amsmath,amssymb,aps,prapplied]{revtex4-2}

\usepackage{graphicx}
\usepackage{dcolumn}
\usepackage{bm}
\usepackage{subfigure}
\usepackage{xcolor}

\begin{document}


\title{{Optimal Design of Dallenbach Absorbers Under Broadband Broad-Angle Illumination}}

\author{Chen Firestein, Amir Shlivinski}
\affiliation{%
 School of Electrical and Computer Engineering, Ben-Gurion University of the Negev, Beer Sheva, 84105, Israel.\\
}%

\author{Yakir Hadad}
\email{yakirhad@tauex.tau.ac.il}
\affiliation{
 School of Electrical Engineering, Tel-Aviv University, Ramat-Aviv, Tel-Aviv, Israel, 69978.\\
}%


\begin{abstract}

{The classical scenario where a \emph{single plane-wave} field impinge a Dallenbach absorber is well studied both theoretically and experimentally. However, occasionally  a \emph{spectrum of plane-waves} impinges the absorber. Such a scenario occurs for example if an antenna is located adjacent to the absorbing layer. In this paper, for this scenario we obtain the absorbing performance bound and design an \emph{optimized layered absorber} that approaches the bound. In a numerical demonstration, we explore a realistic case where a dipole antenna is placed in the vicinity of a finite, electrically thin, Dallenbach absorber backed by a PEC plane in the 6G frequency range. In the absence of the absorbing layer covering the PEC plane, severe scattering from the plane distorts the radiated fields. These distortions are robustly mitigated by the specifically tailored optimal absorber to yield a more desired radiation pattern.  Additionally, we propose a metamaterial realization that emulates the required properties of the absorbing layer for all field polarizations.}

\end{abstract}

\maketitle


\section{Introduction}
Wave absorbers play an important role in wave engineering \cite{ruck1970radar,Sohl2007Physical,Landy2008Perfect,Sohl2008Scattering,Radi2015Thin,Fernandes2019Topological,Krasnok2019Anomalies,Gustafsson2020Upper,Schab2020Trade,qu2022microwave,Abdelrahman2022How}, as they are being used in a variety of electromagnetic applications such as reducing reflections \cite{Ye2013Ultrawideband,cao2022backend,Parameswaran2022ALow,Lin2023Extremely}, enhanced accuracy of antenna measurements in anechoic chambers \cite{Holloway1997Comparison,Chung2003Modeling,Mejean2017Electromagnetic}, decreasing undesired emissions of discrete electric components and printed circuit boards (PCBs) \cite{Arien2015Study,Xing2022Ultra,Khoshniat2021Metamaterial} and to capture light in solar cells \cite{Hagglund2012Plasmonic,Mulla2015Perfect}, to name a few. Among the different types of wave absorbers, Dallenbach layer, a lossy substance backed by a perfect electric conductor (PEC) sheet, is widely used due to its compatibility with planar geometries \cite{ruck1970radar}. For the case of a normal plane wave incidence  a known sum rule \cite{rozanov2000ultimate}  imposes a constraint between the absorption efficacy, thickness and the operating frequency band.  This bound is unavoidably affected by the direction of arrival of the impinging wave and by its polarization. Thus, extensions for oblique incidence have been derived \cite{Doane2013Matching}.

Occasionally the impinging wave is more {complicated} spectrally, involving a {spectrum} of plane waves, arriving at different directions. This is the case, for instance when a source is located adjacent to the absorbing layer, in that case the spectral content comprising the field is much {broader than in} the case of a single plane wave incidence, which is applicable in strictly far-field scenarios.
In this case, the design problem should be formulated by taking into account the {relative contribution to the absorption, i.e. ``absorption significance,'' of different spectral components}. The need to tailor properly the absorption for different spectral components, implies that the absorber has to have a suitably designed non-local behaviour. This can naturally be achieved in a planar layered media \cite{Felsen}.

In the following, we first extend the {canonical} single plane wave absorber bound to the case of spectrally wide field, and include the absorption significance merit in the derived bound. This will be used as our design benchmark.
Later, we verify numerically that the new bound is tight when integrating along the entire {frequency} bandwidth.
Lastly, we set an optimization problem to obtain the optimal design parameters, namely the permittivity $\epsilon$ and electric conductivity $\sigma$ of the layered Dallenbach absorber. While being restricted to the new bound, the optimal solution maximizes the contribution within the desired bandwidth while simultaneously minimizing it outside.
Our approach's applicability is demonstrated through numerical examples involving a half-wavelength dipole antenna positioned above a large, {but} finite, ideal ground plane {(perfectly electric conductor - PEC)}. This configuration leads to the emergence of severe interference pattern in the radiated field, resulting in {some extreme gain variations both frequency and angle dependent that may restrict the use of such physical layout}. {In order to solve this issue and suppress undesired reflections caused by the ground plane, we design a layered Dallenbach absorber in the 6G frequency range, $f\in [7,15] \rm{GHz}$ by taking into account the absorption significance of the antenna's radiation pattern}. {Specifically we consider two distinct cases { when the dipole is placed horizontally}, $(a)$ the hight of dipole antenna above the plane is larger than a wavelength and therefore the radiation pattern suffers from the emergence of grating lobes; while in the second case $(b)$ the dipole antenna is positioned at a hight that is much smaller than a wavelength above the plane and therefore the radiation pattern is not uniform which causes severe gain variations.  {Additionally, to investigate the polarization dependence of the designed absorber, as a third example (c) we replaced the horizontal dipole in case (a) with a vertical dipole}. By designing a specifically optimized absorber, we demonstrate that the severe distortions in the radiation pattern, as seen in cases {$(a)$, $(b)$ or {$(c)$}, can be effectively mitigated, resulting in a more desirable radiation pattern.}}
{Subsequently, we discuss some metamaterial realization aspects for the proposed absorbers that are applied for fabrication with PCB technology.}
%
%

\section{Mathematical Formulation}
\begin{figure}[t]
\centering
    \includegraphics[width=7.9cm]{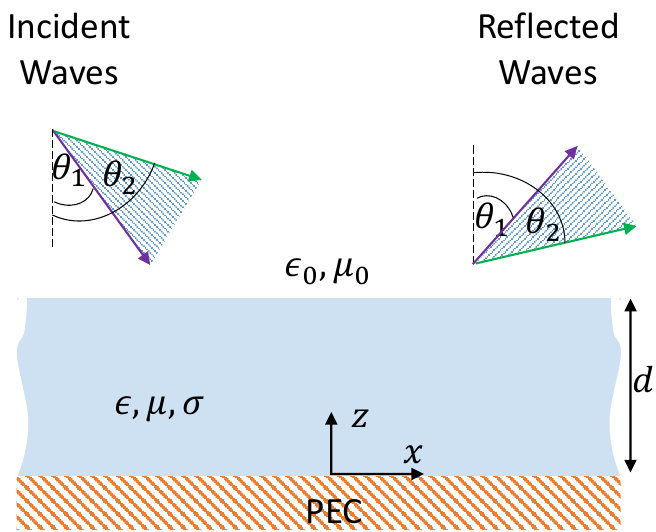}
    \caption{Illustration of the problem. A Dallenbach layer with thickness $d$ is impinged by a wave field that is composed of multiple propagating plane waves within the angular domain $\theta \in \left[\theta_1,\theta_2\right]$.
     As a result of the interaction with the Dallenbach layer, a portion of the wave field is reflected backwards, while the remaining portion is absorbed inside the layer. The wave field may consist of either one or both of the TE and TM polarizations.}
    \label{Model}
\end{figure}

Consider a Dallenbach planar layer consists of a PEC sheet  covered by layered dielectric absorber with electrical permittivity ($\epsilon$), magnetic permeability ($\mu$) and electrical conductivity ($\sigma$)  and surrounded by vacuum ($\epsilon_0,\mu_0$).
The structure extends to infinitely in the $x$ and $y$ directions and has a finite thickness $d$ in the $z$ direction. The absorber is impinged by a wave field that is composed of an angular spectrum of propagating oblique incidence plane waves with either transverse electric (TE) or magnetic (TM) polarizations. See Fig.~\ref{Model} for illustration. Due to the interaction with the Dallenbach layer, some of the wave field is reflected, while the rest is absorbed within the layer. Note that the PEC boundary enforces that there is no transmitted wave beyond the absorbing layer.
Assuming that a single, oblique incidence plane wave with angle $\theta$ impinging the absorber, the following sum-rule is known to bound the absorption efficiency \cite{rozanov2000ultimate,Doane2013Matching},
\begin{equation}
\label{Rozanov_Volakis_Bound}
\left| \int_{0}^{\infty} \!\!\ln \left|\rho(\lambda,\theta)\right|\! \,d\lambda \right|\leq 2\pi^2 \mu_s d
 \begin{cases}
    \rm{cos}\left(\theta\right),& \!\!\rm{TE}\\
   1/\rm{cos}\left(\theta\right),& \!\!\rm{TM}
\end{cases}
\end{equation}
where $\rho$ is the reflection coefficient at the interface between the Dallenbach absorber and its surroundings, $\lambda$ is the free space wavelength and $\mu_s$ is the static relative permeability. In the following, we shall denote the right-hand-side of Eq.~(\ref{Rozanov_Volakis_Bound}) by $\rm{RB} \left(\theta\right)$.
The {expression} in Eq.~\eqref{Rozanov_Volakis_Bound} clearly illustrates the strong dependence of the sum-rule bound on the field polarization as well as on the direction of arrival of the impinging wave.  Therefore, it is clear that Eq.~\eqref{Rozanov_Volakis_Bound} cannot be directly applied for {fields comprising more than a single plane wave, i.e. with a spectrum of plane waves}, such as one expects in the presence of a finite source adjacent to the absorber.

In order to accommodate {for the extended spectral content we propose the following:} First, we multiply both sides of Eq.~\eqref{Rozanov_Volakis_Bound} by a normalized non-negative weight function $W(\theta)$ \cite{Comm1} {that denotes for the ``absorbtion significance'' of the spectral components}. Next, we integrate with respect to the angle of incidence ($\theta$).
In addition, to {accommodate} for real-world applications where systems are typically designed to operate at a specific finite {frequency} bandwidth, we truncate the originally infinite integration wavelength domain into a finite domain $[\lambda_1,\lambda_2]$, where $\lambda_1$ [$\lambda_2$] denote the minimal [maximal] free-space wavelength of interest.  These steps lead to,
\begin{equation}
\label{Relation}
\begin{aligned}
&\int_{\theta_1}^{\theta_2}W(\theta)\left| \int_{\lambda_1}^{\lambda_2} \!\!\ln \left|\rho(\lambda,\theta)\right|\! \,d\lambda \right|d\theta \leq \\
&\int_{\theta_1}^{\theta_2}W(\theta)\left| \int_{0}^{\infty} \!\!\ln \left|\rho(\lambda,\theta)\right|\! \,d\lambda \right|d\theta \leq
\int_{\theta_1}^{\theta_2}W(\theta)\rm{RB}\left(\theta\right) d\theta,
\end{aligned}
\end{equation}
where $[\theta_1,\theta_2]$ are the minimal and maximal desired angles of incidence, respectively and ${\int_{\theta_1}^{\theta_2} W(\theta)d\theta= 1,\hspace{0.1cm} W(\theta)\neq 0 \hspace{0.1cm}\forall \hspace{0.1cm} \theta \in [\theta_1,\theta_2]}$.
The {expression} in Eq.~\eqref{Relation} describes the appropriate {weighted} sum rule (bound) for wave fields composed of a spectrum (multiple) of oblique incidence plane waves.
The right hand side of Eq.~\eqref{Relation} bounds from above the left and middle sides for any selection of the properties of the {layer parameters $\{\epsilon,\sigma\}$ that are embedded in $\rho$}. This gives rise to an optimization problem for designing a Dallenbach absorber.
By employing Eq.~\eqref{Relation} as a $W$-weighted $L^1$ norm on $\theta$ space, we formulate the following optimization problem to obtain the absorber's characteristic parameters,
\begin{equation}
\label{Optimization_problem}
\begin{aligned}
&\min_{\epsilon,\sigma} \int_{\theta_1}^{\theta_2} \left| W(\theta)\rm{RB} -  W(\theta)\left| \int_{\lambda_1}^{\lambda_2} \!\!\ln \left|\rho(\lambda,\theta)\right|\! \,d\lambda \right|  \right| d\theta \\
&\rm{subject\hspace{0.2cm} to} \hspace{1cm} \rho(\lambda,\theta)=\frac{Z_{\mbox {\scriptsize in}}(\lambda,\theta)-\eta_0(\theta)}{Z_{\mbox {\scriptsize in}}(\lambda,\theta)+\eta_0(\theta)}.
\end{aligned}
\end{equation}
The impedance $Z_{\mbox {\scriptsize in}}$ refers to the input impedance at the interface between the absorbing layer and its surroundings, while $\eta_0$ is the free space (TE and TM) characteristic impedance (see \cite{Adler1968Electromagnetic,Collin1966Foundations,Pozar2011Microwave}).
An alternative (equivalent) way to express Eq.~\eqref{Optimization_problem} as a maximization problem is to normalize the cost function by dividing Eq.~\eqref{Relation} by its right-hand side. The resulting value of the expression is between $0$ and $1$, where a score of $1$ indicates the highest achievable value of the cost function, i.e., tightest bound, for any given set of design parameters ($\epsilon,\sigma$). The alternative optimization problem reads,
\begin{equation}
\label{Eqiv_Optimization_problem} 
\begin{aligned}
&\max_{\epsilon,\sigma}\hspace{0.2cm}\frac{\int_{\theta_1}^{\theta_2}W(\theta)\left| \int_{\lambda_1}^{\lambda_2} \ln \left|\rho(\lambda,\theta)\right|\! \,d\lambda \right|d\theta}{\int_{\theta_1}^{\theta_2}W(\theta)\rm{RB} \left(\theta\right) d\theta} \\
&\rm{subject\hspace{0.2cm} to} \hspace{0.4cm} \rho(\lambda,\theta)=\frac{Z_{\mbox {\scriptsize in}}(\lambda,\theta)-\eta_0(\theta)}{Z_{\mbox {\scriptsize in}}(\lambda,\theta)+\eta_0(\theta)}.
\end{aligned}
\end{equation}
%
In the following sections we use Eqs.~\eqref{Relation}-\eqref{Eqiv_Optimization_problem} to explore the tightness of the new sum rule and to design a practical Dallenbach absorber that operates within a finite {but} broad frequency range and wide angular spectrum of incidence waves.

\section{The Tightness of the Sum Rule}\label{tightness_section}
For the sake of concreteness, here, we consider the case of a wave field with an infinite frequency support within the angular domain $[\theta_1,\theta_2]=[0,\pi/3]$. For that purpose we consider a non-magnetic layer ($\mu=\mu_0, \mu_s=1$) with thickness $d=0.4 \left[\rm{m}\right]$. The layer is composed of causal materials satisfying Kramers-Kroning relations, with wavelength dependent permittivity $\epsilon\left(\lambda\right)$ and electric conductivity $\sigma(\lambda)$ having a Lorentzian form \cite{Jackson2007Classical},
%
\begin{equation}
\label{permittivity_conductivity} 
\begin{aligned}
\epsilon\left(\lambda\right)&=\epsilon_0\left(1+\frac{A}{1+j\frac{\lambda_{\rm{rel}}}{\lambda}-(\frac{\lambda_{\rm{res}}}{\lambda})^2}\right) \\
\sigma\left(\lambda\right)&=\frac{\sigma_{\rm{s}}}{1+j\frac{2\pi c_0 \tau_{\sigma}}{\lambda}},
\end{aligned}
\end{equation}
where $A$ denotes the strength of the dielectric response, $\lambda_{\rm{res}} [\lambda_{\rm{rel}}]$ is the resonant [relaxation] wavelength, $\sigma_{\rm{s}}$ is the static (long wavelength) electric conductivity, $c_0=1/\sqrt{\epsilon_0\mu_0}$ is the speed of light in vacuum and $\tau_\sigma$ is a time constant of the conductor.
By combining Ampere's law with Ohm's law and Eq.~\eqref{permittivity_conductivity}, an effective electric permittivity, $\epsilon_{\rm{eff}}(\lambda)$, can be expressed. This effective permittivity takes into account both material polarization and conductance effects and is given by
\begin{equation}\label{RF_lossy_disp}
\epsilon_{\rm{eff}}(\lambda)=\epsilon\left(\lambda\right)+\frac{\sigma\left(\lambda\right)}{j2\pi c_0\lambda^{-1}}.
\end{equation}

\begin{figure}[!h]
\centering
   \subfigure[]{%
     \includegraphics[width=0.49\columnwidth]{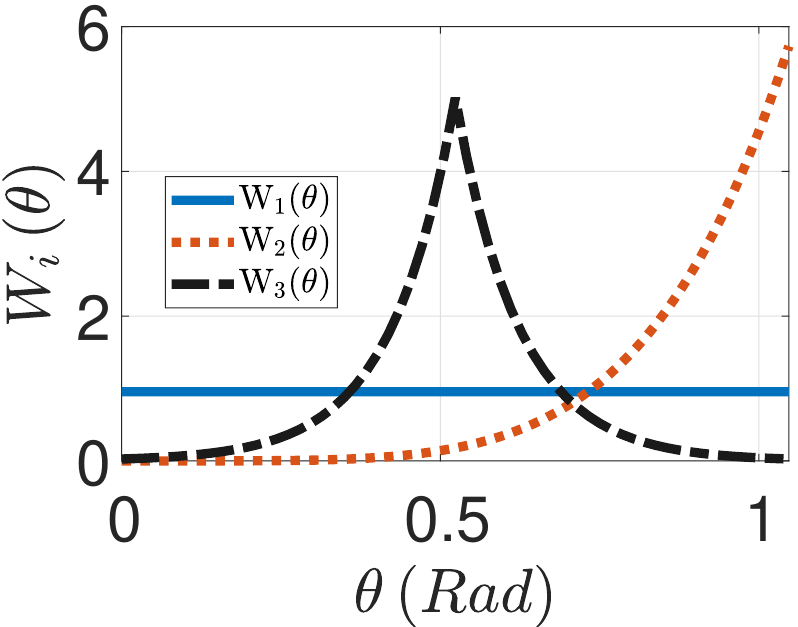}}
\\
   \subfigure[]{%
     \includegraphics[width=0.49\columnwidth]{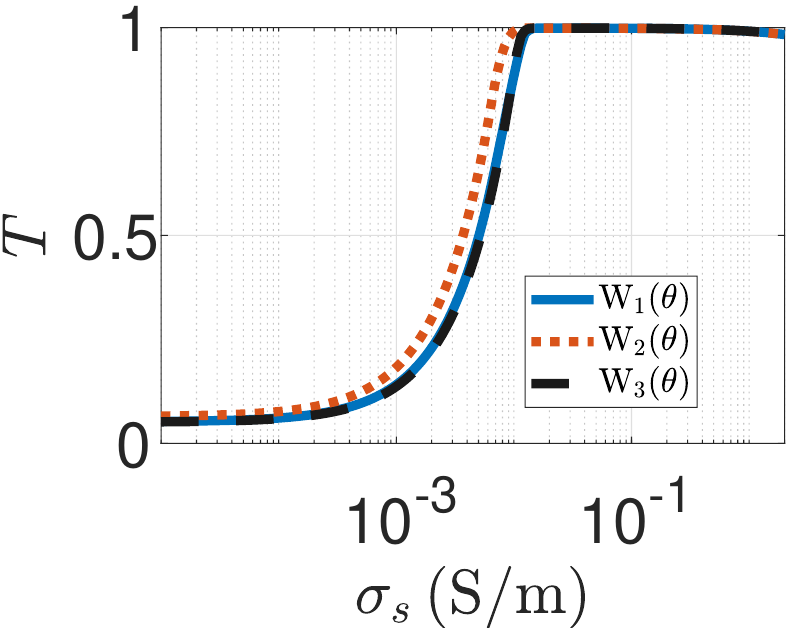}}
    \subfigure[]{%
     \includegraphics[width=0.49\columnwidth]{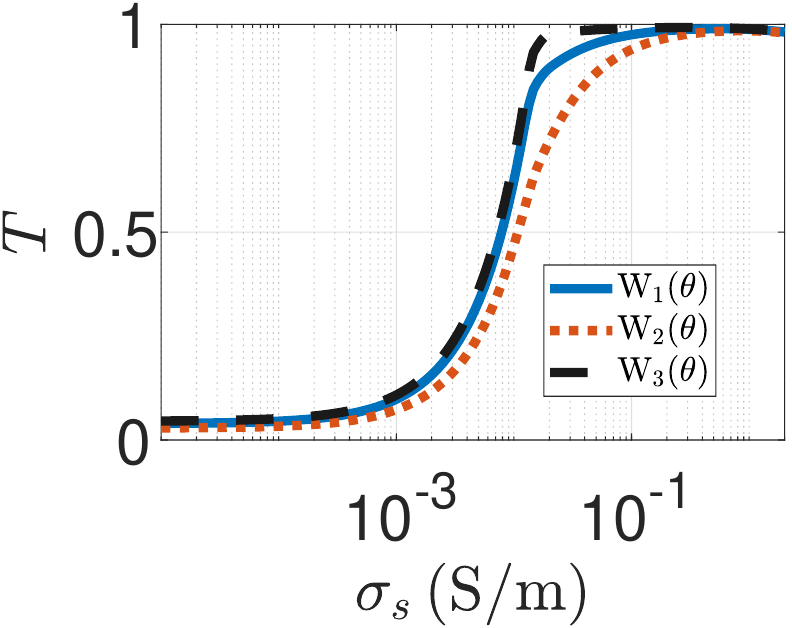}}
\caption{Tightness of the bound for several weight functions. In accordance with Eq.~\eqref{Relation}, the ratio is bounded from above by $1$ and is non-negative. The ratio is calculated for a wide range of $\sigma_{\rm{s}}$, while holding all other parameters in Eq.~\eqref{permittivity_conductivity} constant. (a) The three weight functions. (b) TE polarization. (c) TM polarization.}
\label{FullWaveband_Bound}
\end{figure}

For the sake of concreteness, we set the following parameters: $A=1, \lambda_{\rm{res}}= \lambda_{\rm{rel}} = 1\left[\rm{m}\right], \tau_{\sigma}=10^{-12} \left[\rm{sec}\right]$, while the only remaining free-parameter is $\sigma_{\rm{s}}$ that will be scanned over a wide range of values to demonstrate its effect on the tightness of the bound.
In addition, we define three {typical} weight functions \{$W_1(\theta),W_2(\theta),W_3(\theta)$\} within the angular domain $\theta \in [0,\pi/3]$ (see Fig.~\ref{FullWaveband_Bound} (a)),
\begin{equation}
\label{Weight functions} 
\begin{aligned}
&W_1(\theta)= \frac{3}{\pi},\\[1ex] 
&W_2(\theta)= 4.5496\hspace{0.1cm} \theta^5 \\[1ex]
&W_3(\theta)= 5.015\hspace{0.1cm} e^{-10|\theta-\pi/6|},\\
\end{aligned}
\end{equation}
and zero elsewhere. The three weight functions (recall, these enforce the ``absorption significance'' of each spectral component) were selected to have specific characteristics. Firstly, $W_1(\theta)$ is uniformly distributed, indicating that all angles hold equal importance. Secondly, $W_2(\theta)$ prioritizes larger angles, where extreme values are observed in the sum rule (see Eqs.~\eqref{Rozanov_Volakis_Bound} and \eqref{Relation}). Lastly, $W_3(\theta)$ exhibits symmetry around $\theta=\pi/6$ and follows an exponential distribution.
We define the \emph{tightness} of the bound as the ratio in the argument of the optimization problem in Eq.~\eqref{Eqiv_Optimization_problem},
\begin{equation}
\label{Ratio}
T=\frac{\int_{\theta_1}^{\theta_2}W_i(\theta)\left| \int_{0}^{\infty} \ln \left|\rho(\lambda,\theta)\right|\! \,d\lambda \right|d\theta}{\int_{\theta_1}^{\theta_2}W_i(\theta)\rm{RB} \left(\theta\right) d\theta},\qquad 0 \le T \le 1.
\end{equation}
Figures~\ref{FullWaveband_Bound} (b) and (c) depict the evaluation of $T$ for multiple values of $\sigma_{\rm{s}}$  and for each $W_i(\theta)$ with $i\in {1,2,3}$ and for both TE and TM polarizations. The integral in the numerator should be performed over an infinite wavelength range. This is obviously impossible numerically. Instead, the integration is performed over {an extended but finite wavelength range} $\lambda \in \left[0.001,10^4\right] \left[\rm{m}\right]$ and converges to the infinite integration result is verified. The evaluation of the integration in the numerator uses the exact value of the  reflection coefficient, $\rho(\lambda, \theta)$, that was calculated using the expression given in Eqs.~\eqref{Optimization_problem}-\eqref{Eqiv_Optimization_problem} where the impedance $Z_{\rm in}$ is given by the transmission-line model of the layered media \cite{Adler1968Electromagnetic,Collin1966Foundations}. A tight bound with values approaching $1$, as shown in Fig.~\ref{FullWaveband_Bound}(b,c), can be observed for any of the three weight functions considered, provided that the free parameter $\sigma_{\rm{s}}$ is properly selected.
%
It is important to note that the scanning process {for finding optimal parameters} can be further {fine-tuned} by including additional free-parameters, such as $A$, $\lambda_{\rm{res}}$, $\lambda_{\rm{rel}}$, and $\tau_{\sigma}$ or even to incorporate multiple resonance terms in the Lorentizian model. By exploring different combinations of these parameters, it is possible to identify alternative sets that may also maximize the ratio.

\section{Optimal Absorber Design}\label{Opt_Approach}
\begin{figure}[!h]
\centering
    \includegraphics[width=8.0cm]{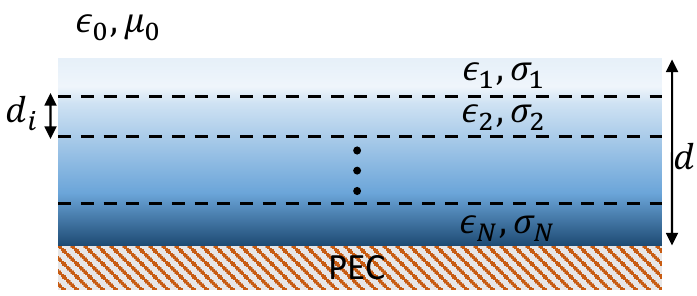}
    \caption{A piecewise homogenous design where the non-magnetic absorbing layer of thickness $d$ is truncated into $N$ parallel sub-layers with identical thickness $d_i=d/N$ ($i \in [1,N]$), each of which is homogeneous with $\epsilon_i,\sigma_i$.}
    \label{Sub-layers}
\end{figure}

In practical systems, absorption performance is typically optimized within a specific bandwidth $\lambda \in [\lambda_1,\lambda_2]$, $\lambda_1>0$ and $\lambda_2 < \infty$. In light of the absorption efficiency sum-rule, it is clear that to maximize absorption within a certain frequency range it is essential to minimize it elsewhere. The optimization approach developed here is formulated to achieve this goal. Because of the design's field polarization sensitivity, the optimization needs to be conducted separately for TE and TM polarizations.
The optimization process that we consider allows also for the design of a stratified inhomogeneous Dallenbach layers. In that case, we consider a piecewise homogeneous design, where, the layer of thickness $d$ is truncated into $N$ parallel sub-layers with identical thickness, each of which is homogeneous (see Fig.~\ref{Sub-layers} for illustration).
For any value of $N$ we aim to find optimal design parameters ($\epsilon ,\sigma$) that maximizes the tightness, $T$, in Eq.~\eqref{Ratio}. For each sub-layer, we consider the following two models for the design parameters:\\
\begin{itemize}
\item{Model $1$ - constant parameters.} We use Eq.~\eqref{permittivity_conductivity} with $\lambda_{\rm{rel}}=\lambda_{\rm{res}}\approx0$ and $\frac{2\pi c_0 \tau_{\sigma}}{\lambda}\ll 1$, resulting in an approximately constant $\epsilon=\epsilon_0\epsilon_r$ and $\sigma=\sigma_{\rm{s}}$, where $\epsilon_r=1+A$ is the relative permittivity. In this model, there are two design parameters, $\epsilon_r$ and $\sigma$, which provide only a conductive loss mechanism.
\item{Model $2$ - frequency dependent parameters.} We use Eq.~\eqref{permittivity_conductivity} with $\epsilon(\lambda)$ providing three design parameters, $A$, $\lambda_{\rm{rel}}$, and $\lambda_{\rm{res}}$. To reduce the number of design parameters, it is convenient to set $\sigma_s=0$, providing only a polarization loss mechanism.
\end{itemize}
Model $1$ is widely used in microwave engineering, such as for example in transmission-lines where the parameters usually vary slowly within the desired frequency band to maintain wideband impedance matching. In this model the material resonances are typically located at much higher frequencies than these of interest, as opposed to Model $2$ where they are within (or near the edge) of the operational region. Both models satisfy Kramers-Kronig relation over the entire infinite frequency range.
Note that other models, such as incorporating electric conductivity $\sigma(\lambda)$ in Model $2$ or incorporating multi-resonance polarization response, may also be applicable. However, as the complexity of the model increases, with additional design parameters per sub-layer, the optimization process becomes more involved while the relative improvement decreases.
The optimization problem to obtain the absorber's parameters giving the tightest design is \emph{non-convex}.  To solve this optimization problem, we adopt a four-level search approach. First forming a relatively coarse calculation grid. In Model 1 realization, the parameters $\epsilon_r$ and $\sigma$ of each layer are uniformly distributed with $M$ points in the rages $\epsilon_{r,\rm{min}} \le \epsilon_r \le \epsilon_{r,\rm{max}}$ and $0 \le \sigma \le \sigma_{\rm{max}} \left[\rm{S/m}\right] $, for some parameters  $\{\epsilon_{r,\rm{min}},\epsilon_{r,\rm{max}},\sigma_{\rm{max}}\}$.  While in  Model $2$ realization, the parameters of each layer $A$, $\lambda_{\rm{rel}}$ and $\lambda_{\rm{res}}$ are uniformly distributed with $M$ points in the ranges $A_{\rm{min}} \le A \le A_{\rm{max}}$ and $\lambda_{\rm{min}} \le \lambda_{\rm{rel}}, \lambda_{\rm{res}} \le \lambda_{\rm{max}} \left[\rm{m}\right]$ for some parameters  $\{A_{\rm{min}},A_{\rm{max}},\lambda_{\rm{min}},\lambda_{\rm{max}}\}$. Note that the lower and upper values of the grid may vary based on the requirements, such as frequency bandwidth, angular domain, etc. Overall, with the parameters of $N$ sub-layers, the total number of calculation grid points in Model $1$ is $M^{2N}$, and for Model $2$ is $M^{3N}$. To ensure memory limitations are not exceeded, for a given $N$, we choose $M$ such to limit the number of grid points to no more than $500$ million. 
For each calculation grid point, in the first calculation level, we evaluate the cost function in Eq.~\eqref{Optimization_problem} (or its equivalent in Eq.~\eqref{Eqiv_Optimization_problem}).
In the second step, we focus on the top-performing results from the first-level, coarse grid, calculation step, limiting our search to no more than {$1\%$} of the total number of grid points. To refine our search further, we perform a local search around each candidate solution of the first-level, evaluating the cost function at steps that are smaller by factor $10$ than in the first-level. Once we have explored possible refinement in the vicinity of the top-performing results of the previous step, the most optimal solution obtained so far is set as an input to the third step. In the third step, we employ a subsequent round of local search, which involves exploring the immediate surroundings of the optimal solution using smaller step sizes, specifically reduced by a factor of $25$ compared to the initial step. This process allows us to fine-tune the optimal solution and achieve even greater refinement. As a final step we perform the optimization procedure described in steps $1-3$ three additional times, adjusting arbitrarily the parameters   $\{\epsilon_{r,\rm{min}},\epsilon_{r,\rm{max}},\sigma_{\rm{max}},A_{\rm{min}},A_{\rm{max}},\lambda_{\rm{min}},\lambda_{\rm{max}}\}$ in the first step. This step aims to improve the optimized results, as the initial grid points are modified. The entire process is done for both TE and TM polarizations and for several values of $N$.
In the following section, we illustrate the design procedure for a metamaterial structure to emulate the necessary effective parameters derived from the optimization process. For this purpose, employing Model $1$ is more practical due to its more feasible requirements in the quasistatic regime.

\section{Practical numerical example - metamaterial absorber for 6G applications}
In this section we demonstrate {by} numerical examples how {the absorber's design introduced in the previous sections helps} to overcome {fundamental issues} with antennas that are placed {in the vicinity of a large but \emph{finite}} reflective surfaces (modelled as PEC ground plane).

\subsection{{Horizontal dipole above an optimized Dallenbach layer}}
{We start by noting that the H-plane radiation pattern of a half-wavelength dipole which is inherently omnidirectional in free space}, exhibits variable radiation performance when parallel to a large ground plane \cite{Balanis2016Antenna}. {This comes due to a reflected field from the ground plane that can be interpreted as a contribution of an image source that is anti-symmetrically located beyond the PEC. Note that this intuitive description neglects edge diffraction effects resulting from the finite nature of the ground plane.}

\begin{figure}[!h]
\centering
   \subfigure[]{%
     \includegraphics[width=0.56\columnwidth]{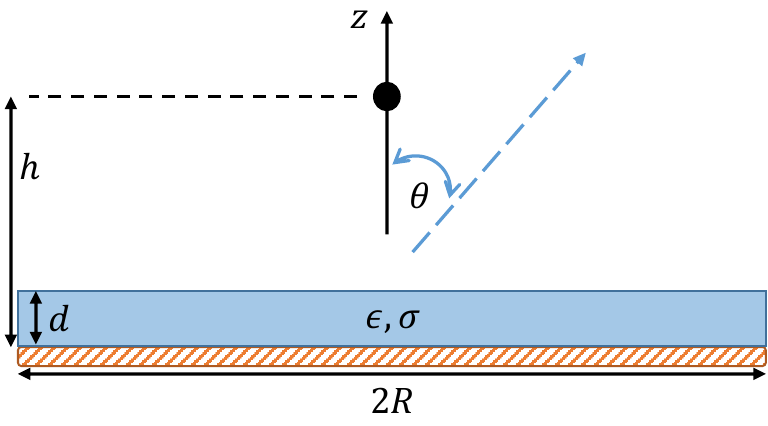}}
    \subfigure[]{%
     \includegraphics[width=0.42\columnwidth]{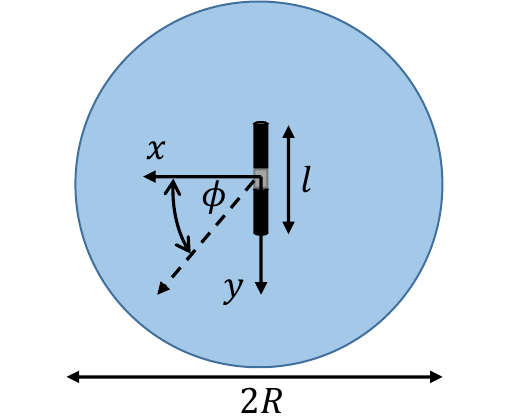}}
\\
   \subfigure[]{%
     \includegraphics[width=0.49\columnwidth]{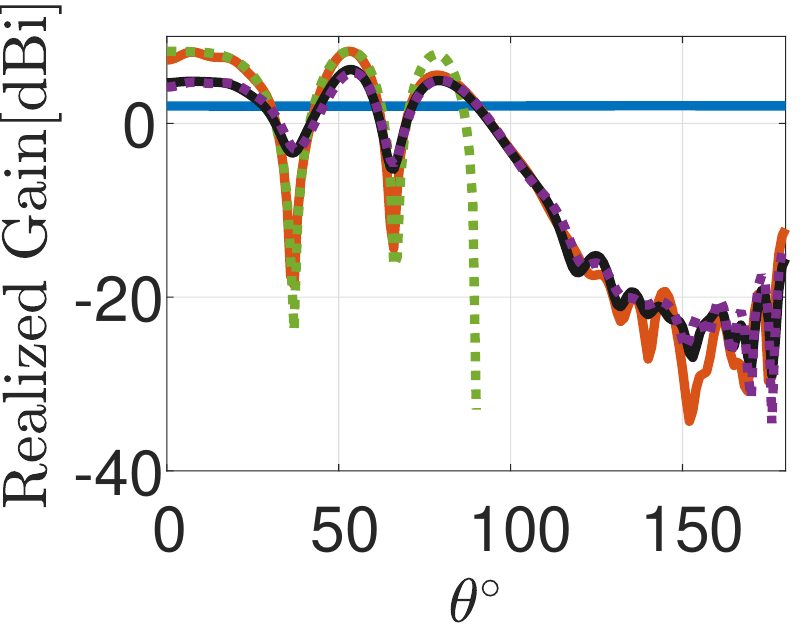}}
    \subfigure[]{%
     \includegraphics[width=0.49\columnwidth]{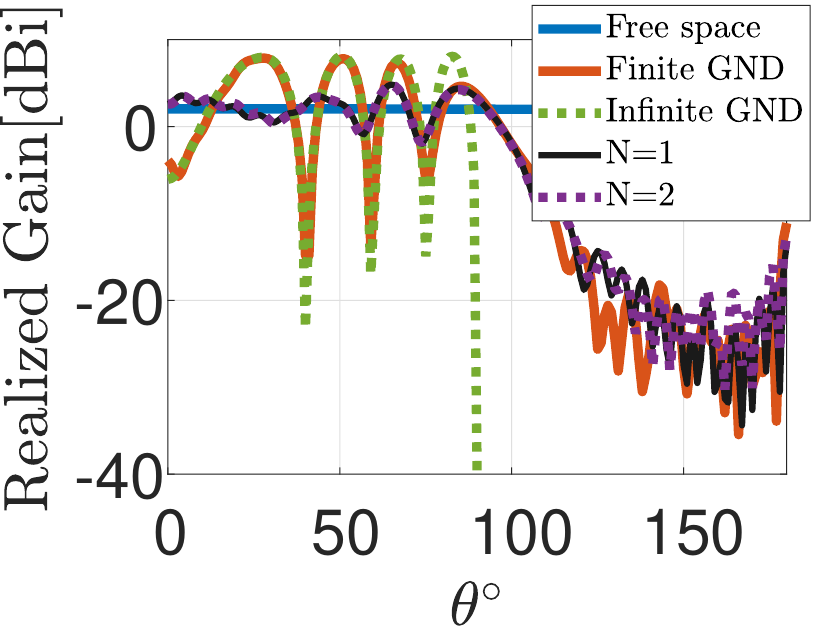}}
\caption{Problem illustration. A half-wavelength dipole is placed above, in parallel to a Dallenbach absorber {($h=0.0536[\rm{m}]$, corresponding to  $1.25\lambda_{\rm{max}}$)}. (a) Side view. (b) Top view. The gray rectangle illustrates a lumped port that is set to excite the structure. (c) and (d) present the H-plane realized gain obtained from HFSS for {$f=7 \rm{GHz}$} and {$f=11 \rm{GHz}$}, respectively. Radiation in free space (blue), the radiating element in the presence of an infinite ground plane (dotted green) and finite ground plane (dark red), the element in the presence of an optimal Dallenbach absorber with $N=1$ (black), and $N=2$ (dotted purple).}
\label{Structure_illust}
\end{figure}

In {the first example}, the dipole is positioned at $h=1.25\lambda_{\rm{max}}$ above a circular finite ground plane of radius of $R=5\lambda_{\rm{max}}$ ({$\lambda_{\rm{max}}=0.0429 [\rm{m}]$, corresponding to the free space wavelength at $f=7 [\rm{GHz}]$}).
{The non-negligible  distance however $>\lambda$ (in the band) between the dipole and the ground plane induces the presence of grating lobes that severely distort the radiation field by introducing nulls in the field radiation pattern}. {To address this issue and mitigate the impact of the distorting reflected wave}, we aim to optimize a Dallenbach absorber with a thickness of {$d=0.005 [\rm{m}]$} in the {6G} frequency range $f\in[7,15] \rm{GHz}$. {Physical layout of the system} is depicted in Figs.~\ref{Structure_illust}(a)--(b), {showing} side and top views, respectively. It is important to note that as the frequency increases, more grating lobes emerge while the distance $h$ between the radiating element and the ground plane remains constant. We {use a cosine} weight function, {$W(\theta)=\rm{cos}(\theta)$}, over $\theta \in [0,\pi/2]$ {to denote the absorption significance.} Given the predominant TE polarization aligned with the wire in the far-field of the half-wavelength dipole antenna, we {apply} the optimization approach, detailed in Sec.~\ref{Opt_Approach} { with  one or two absorbing layers ($N=1,2$)} to determine optimal values of $\epsilon_r$ and $\sigma$ {for suppressing TE polarized reflected} waves.
{The optimized values are} : {$\epsilon_r=2.807$, $\sigma=0.996 [\rm{S/m}]$} for $N=1$ (resulting in a score of {0.4213} in Eq.~\ref{Eqiv_Optimization_problem}), and {$\epsilon_r=\{3.7627,5.7119\}$, $\sigma=\{0,5.2881\} [\rm{S/m}]$} for $N=2$ (yielding a score of {0.4855} in Eq.~\ref{Eqiv_Optimization_problem}). These results were obtained with the grid constraints $\epsilon_{r,\rm{min}}=1$, $\epsilon_{r,\rm{max}}=5$, $\sigma_{\rm{max}}=1 [\rm{S/m}]$ (for $N=1$), and $\epsilon_{r,\rm{min}}=2$, $\epsilon_{r,\rm{max}}=7$, $\sigma_{\rm{max}}=6 [\rm{S/m}]$ (for $N=2$).
To assess the system's performance, both with and without the absorbing layer, a full-wave electromagnetic simulation was carried out using the High-Frequency Structure Simulator (HFSS) \cite{HFSS}.
A lumped port was employed to excite the half-wavelength dipole. {Given the narrow frequency band nature of the radiating element which is in contrast with the extended range of the 6G frequency band, a scaled redesign was applied to match the element's length ($l$) with the corresponding free space wavelength ($l\sim \lambda/2$) at each frequency}. This involved varying the length of the radiating element to maintain the desired proportional relationship with the frequency.
Figures \ref{Structure_illust} (c) and (d) display the numerically obtained realized gain (H-plane) from HFSS for {$f=7 \rm{GHz}$ and $f=11 \rm{GHz}$}, respectively. Each figure features 5 lines representing the realized gain for different scenarios: free space radiation of the half-wavelength dipole only (blue), the dipole in the presence of an infinite ground plane (dotted green), the dipole in the presence of a finite ground plane (dark red), the dipole in the presence of an optimal Dallenbach absorber with $N=1$ (black), and $N=2$ (dotted purple).
As seen in Figs.~\ref{Structure_illust} (c) and (d), the lack of an absorbing layer results in multiple angles where the realized gain is nullified. This is due to destructive interference between the direct ray and the reflections caused by the finite ground plane. {In contrast, as we introduce the electrically thin optimized Dallenbach absorber, the severe effect of the reflected waves from the ground plane, causing the nulls, is mitigated in such a manner as to nearly eliminating any trace of the nulls at higher frequencies further into the band to give antenna's performance that is closer to what is expected in free space.} 
{To explore and assess the effect of the absorber on the dipole's radiation pattern, we introduce a flatness parameter ($F$) as follows,
\begin{equation}
\label{Flatness}
F\!\!\triangleq \!\!\left [\frac{1}{2\pi}\!\!\int_{0}^{\pi/2} \!\!\left[\frac{G(\theta)}{G_{\rm{FSD}}(\theta)}+\frac{G_{\rm{FSD}}(\theta)}{G(\theta)}\right]^{2} \!\!d\theta \right ]^{\frac{1}{2}}, \quad F \ge 1
\end{equation}
where $G_{\rm{FSD}}$ is the realized gain of the free-space dipole and $G$ represents the realized gain under examination for finite GND, infinite GND and optimal absorber designs. Applying Eq.~\eqref{Flatness} with the simulation results in Fig.~\ref{Structure_illust} (c) reveals that without the absorber (finite GND) $F=8.359$. With the absorber $F=1.2977$ for $N=1$ and $F=1.2528$ for $N=2$, indicating an improvement of about $84.7\%$. Similarly, in Fig.~\ref{Structure_illust} (d) without the absorber (finite GND) $F=4.5312$. With the absorber $F=1.0686$ for $N=1$ and $F=1.0659$ for $N=2$, indicating an improvement of about $76.4\%$. }
%
%
Similar results to those shown in Figs.~\ref{Structure_illust} (c) and (d) were achieved at frequencies of {$f=9,13,15 [\rm{GHz}]$}.

\begin{figure}[!h]
\centering
   \subfigure[]{%
     \includegraphics[width=0.49\columnwidth]{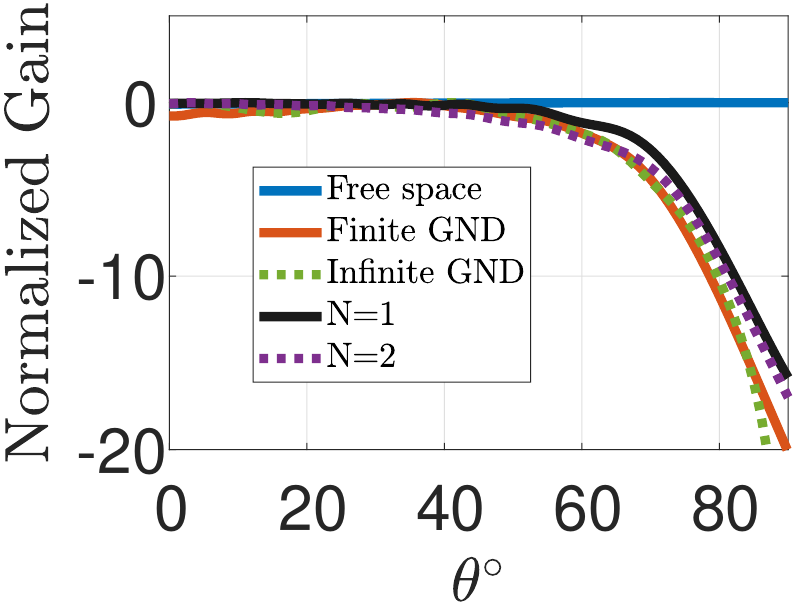}}
    \subfigure[]{%
     \includegraphics[width=0.49\columnwidth]{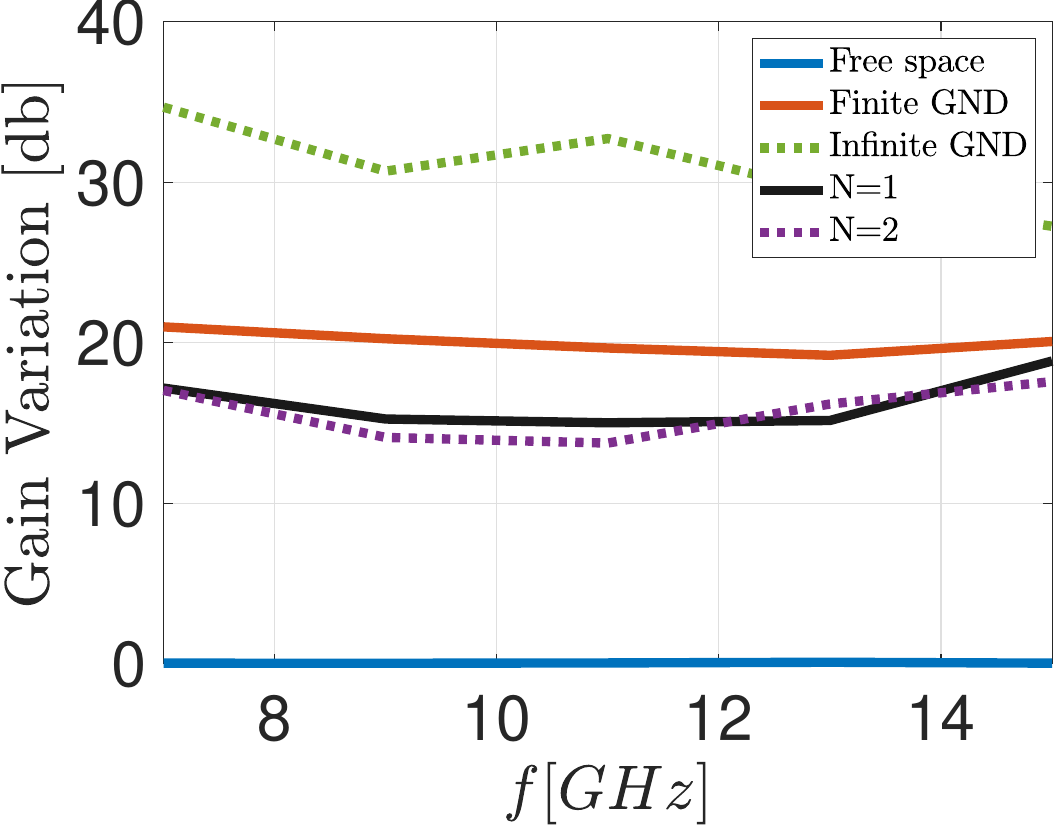}}
\caption{ {(a) A typical normalized gain radiation pattern (in db) at $f=13 \rm{GHz}$ with $h=0.007[\rm{m}]$, corresponding to $h/\lambda=0.3033$ for the five cases (as in Fig. ~\ref{Structure_illust}). (b) depicts the gain variation in the upper radiation hemisphere at the different frequencies in the 6G band. It is readily noted the for both absorbers' realizations ($N=1,2$, solid-black and dotted-purple lines, respectively). In both figures the blue line represents the free space case of the half-wavelength dipole, the dotted green line shows the dipole in the presence of an infinite ground plane, the dark red line indicates the dipole with a finite ground plane, and the black and dotted purple lines depict the dipole in the presence of an optimal Dallenbach absorber with  $N=1$ and $N=2$, respectively.
}}
\label{Structure_illust2}
\end{figure}

{In the first example, we examined the case where the radiating element was positioned more than $1.25\lambda$ away from the absorber, which was utilized to demonstrate the reduction of grating lobes that caused nulls to appear in the radiation pattern in specific directions.
Next, as a second example, we explore whether the \emph{same} absorber that was designed in the first example, above, can effectively reduce reflections when the dipole is placed in the immediate vicinity of the surface ($h\ll\lambda$). In this case, grating lobes do not occur. However, the gain varies rapidly with $\theta$, which is \emph{undesirable} for quasi-omnidirectional and quasi-isotropic antennas \cite{Bybi2008Aquasi,Shah2021Recent,Zhang2022LowProfile}. A common performance metric for such antennas is ``gain variation'', defined as the difference between the maximum and minimum gain the angular band measured in decibels . In our scenario, the large PEC ground plane enforces zero radiation at grazing angles ($\theta\sim\pi/2$) and in the blocked lower hemisphere  ($\theta\in[\pi/2,\pi]$) and therefore inducing some gain variation in upper radiation hemisphere. An efficient wave absorber may significantly reduce the gain variation in the upper hemisphere. {This is not a trivial task at all, as omni-directional and isotropic antennas are typically small compared to the operating wavelength and their radiation pattern is sensitive when placed in the vicinity of large metallic reflectors. To that end, Fig.~\ref{Structure_illust2}(a) depicts a typical normalized gain radiation pattern (in db) at $f=13 \rm{GHz}$ with $h=0.007[\rm{m}]$, corresponding to $h/\lambda=0.3033$ for the five cases (as in Fig. ~\ref{Structure_illust}) similar patterns are obtained for different frequencies in the band. It should be noted that due to the close proximity of the antenna to the finite ground plane (dark red line) and the strong electromagnetic coupling, the maximum of the gain is shifted from the boresight (z axis) direction and increase the gain variation. Figure~\ref{Structure_illust2}(b) depicts the gain variation in the upper radiation hemisphere at the different frequencies in the 6G band. It is readily noted the for both absorbers' realizations ($N=1,2$, solid-black and dotted-purple lines, respectively) the gain variation is smaller by at least 4 [db] along nearly all the 6G band with respect to the finite ground plane with no absorber case (solid dark red line). Note that smaller gain variation imply for relatively more uniform angular pattern, therefore the use of a Dallenbch absorber is favorable as it provide better results as to the no-absorber case. It should be emphasized that in a free-space case, where the pattern is omni-directional, the gain variation is 0 [db]. This is the most desired case however due to the practically constrained physical space for the construction of the optimal absorber a \emph{complete} mitigation of the gain variation is limited however this example indicate that ``correction'' in the gain variation is indeed feasible.}}


\subsection{{Vertical dipole above an optimized Dallenbach layer}}
\begin{figure}[!h]
\centering
   \subfigure[]{%
     \includegraphics[width=0.56\columnwidth]{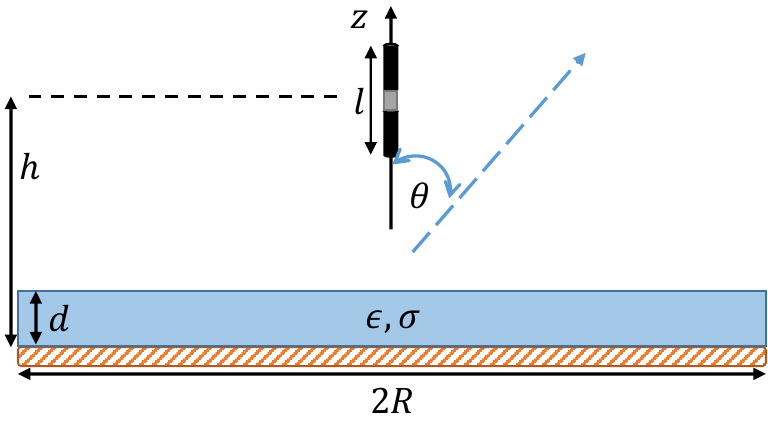}}
    \subfigure[]{%
     \includegraphics[width=0.42\columnwidth]{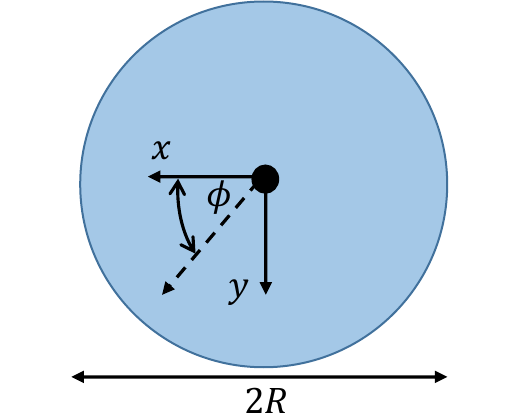}}
\\
   \subfigure[]{%
     \includegraphics[width=0.49\columnwidth]{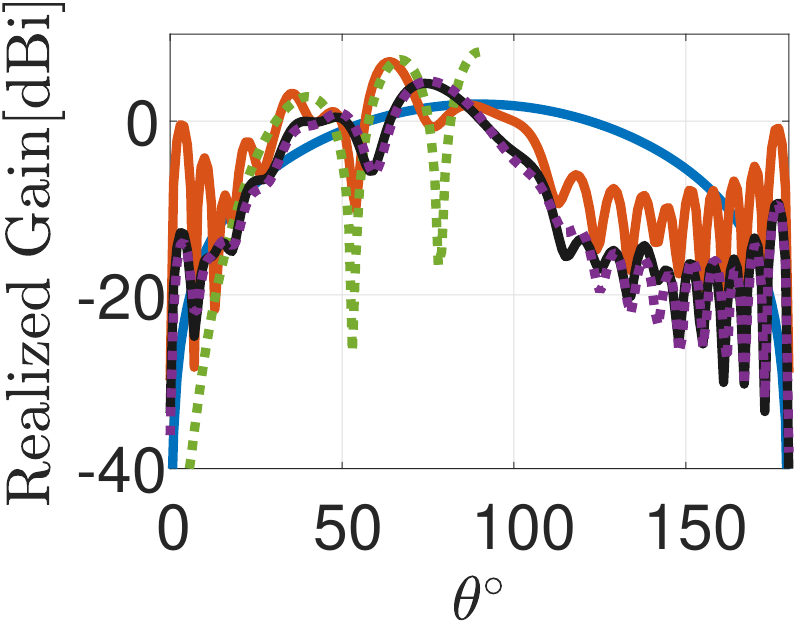}}
    \subfigure[]{%
     \includegraphics[width=0.49\columnwidth]{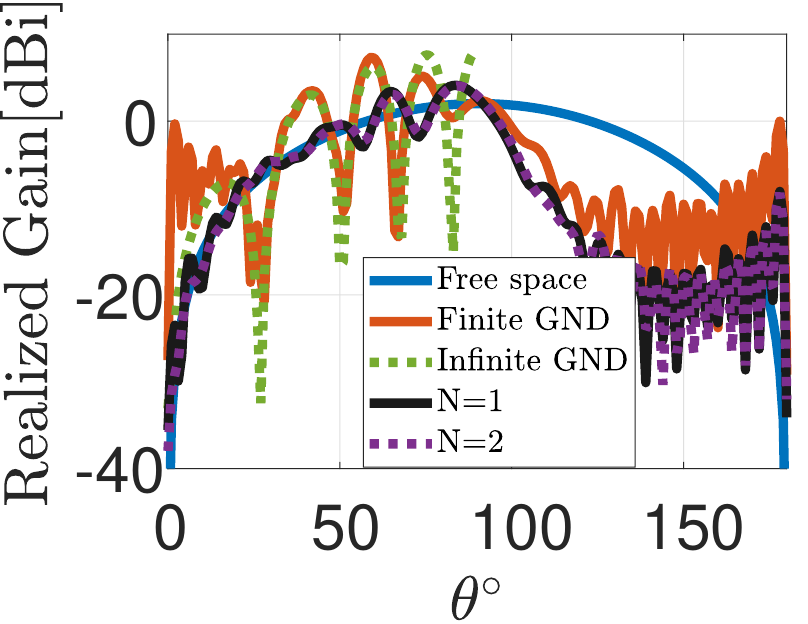}}
\caption{{A half-wavelength dipole is placed above (vertically oriented) to a Dallenbach absorber ($h=0.0536[\rm{m}]$, corresponding to  $1.25\lambda_{\rm{max}}$). (a) Side view. (b) Top view. (c) and (d) present the realized gain obtained from HFSS for $f=7 \rm{GHz}$ and $f=11 \rm{GHz}$, respectively (in the Z-X plane). Radiation in free space (blue), the radiating element in the presence of an infinite ground plane (dotted green) and finite ground plane (dark red), the element in the presence of an optimal Dallenbach absorber with $N=1$ (black), and $N=2$ (dotted purple).}}
\label{Structure_illust_vert}
\end{figure}

{Here, we aim to verify whether the \emph{previously designed} absorber can reduce reflections caused by a half-wavelength vertical dipole placed at a height of $h = 1.25\lambda_{\rm{max}}$ above the same circular finite ground plane. It is important to note that at this stage, no further optimization steps are performed. Instead, we use the previously optimized parameters:  $\epsilon_r=2.807$, $\sigma=0.996 [\rm{S/m}]$ for $N=1$  and $\epsilon_r=\{3.7627,5.7119\}$, $\sigma=\{0,5.2881\} [\rm{S/m}]$ for $N=2$.
Figures~\ref{Structure_illust_vert} (a) and (b) show the side view and top view of the geometry, respectively.
Figures \ref{Structure_illust} (c) and (d) show the numerically obtained realized gain (Z-X plane) from HFSS at frequencies of $f = 7 , \rm{GHz}$ and $f = 11 , \rm{GHz}$, respectively. Each figure includes five lines representing the realized gain for various scenarios, as described previously: free-space radiation of the half-wavelength dipole only (blue), the dipole in the presence of an infinite ground plane (dotted green), the dipole in the presence of a finite ground plane (dark red), the dipole in the presence of an optimal Dallenbach absorber with $N=1$ (black), and $N=2$ (dotted purple).
It is evident that the absorber effectively mitigates reflections, as the realized gain in the presence of the absorber closely matches the gain observed when the dipole is placed in free space. To quantify this effect, the $F$ parameter, defined in Eq.~\eqref{Flatness}, is used.
In Fig.~\ref{Structure_illust} (c) without the absorber (finite GND) $F=53.09$. With the absorber $F=3.8884$ for $N=1$ and $F=2.4733$ for $N=2$, indicating an improvement of  $92.67\%$ and $95.34\%$, respectively. Similarly, in Fig.~\ref{Structure_illust} (d) without the absorber (finite GND) $F=78.8501$. With the absorber $F=3.1208$ for $N=1$ and $F=1.9654$ for $N=2$, indicating an improvement of  $96.04\%$ and $97.5\%$, respectively.
Similar results to those presented in Figures~\ref{Structure_illust_vert} (c) and (d) were obtained at frequencies of  $f=9,13,15 [\rm{GHz}]$. }

\section{Metamaterial realization of the optimized layer parameters}
Our design, as discussed in the previous section, is aimed for the microwave range where the proper lossy dielectric dispersion is dictated by Model 1 that was introduced in Sec.~\ref{Opt_Approach}. Thus, we seek to realize the desired complex relative permittivity  (see Eq.~(\ref{RF_lossy_disp}))
\begin{equation}
\label{eps_desired}
\epsilon_{\rm{desired}}(f)=\epsilon_r+\sigma/(j2\pi f \epsilon_0).
\end{equation}
This design challenge is addressed in this section, where we provide composite metamaterial layer that emulated effectively the desired effective parameters ($\epsilon_r, \sigma$) obtained through the optimization carried out in the previous section.
{This design should achieve the desired permittivity profile given in Eq.~\eqref{eps_desired} for {TE and TM waves} to incorporate with the broad spectral range and the two dipole orientations, horizontal and vertical with respect to the absorbing layer.} To achieve this goal, we propose a method for designing a metamaterial structure with minimal frequency and spatial dispersion within the quasistatic regime. We {start} with a 2D conductive wired metamaterial structure, where the effective permittivity was previously defined in \cite{Tretyakov2003Analytical}. In our recent work \cite{firestein2021absorption}, we revisit the calculations outlined in \cite{Tretyakov2003Analytical}, specifically for normal incidence waves. This involves the introduction of a resistor ($R'$) connecting the two arms of the wire within each unit cell, embedded in a host material with relative permittivity $\epsilon_{h}$, resulting in an effective relative permittivity $\epsilon_{\rm{analytical}}$,
\begin{equation}
\label{eps_h}
\epsilon_{\rm{analytical}}(f)=\epsilon_{h}-\frac{1}{\epsilon_{0}(2\pi f)^{2}a^{2}\tilde{L}-j\epsilon_{0}(2\pi f) a^{2}\tilde{R}},
\end{equation}
where {$a=1 [\rm{mm}]$} represents the dimension of the cubical unit cell, $r_{0}=0.15 [\rm{mm}]$ is the wire radius, $\tilde{L}=\mu_{0}/(2\pi) \ln \left[a^{2}/(4r_{0}(a-r_{0}))\right]$ and $\tilde{R}=R'/a  [\rm{\Omega/m}]$ are the inductance and resistance per unit length, respectively.
Embedding loaded conductive wires in a host material can pose implementation challenges. Hence, we leverage PCB technology \cite{Abdul2002Simple,Jones2004PCB} by substituting the wires with flat strips, each having a width of $2r_{0}$. The resistors can be implemented in PCB technology as embedded resistors, such as for example \cite{EmbeddedRes}. The host material serves as the substrate for the PCB.
However, substituting the metamaterial loaded conductive wires embedded in a host material with a PCB is non-trivial, and the design process involves the following steps.
\subsection{{Design procedure}}
Step 1 - Fitting to Eq.~(\ref{eps_h}). A parametric sweep was conducted for the wired metamaterial on $\epsilon_h$ and $R'$ to minimize the root mean squared error (RMSE) between $\epsilon_{\rm{desired}}$ and $\epsilon_{\rm{analytical}}$. The optimal values were found to be {$\epsilon_h=2.9,R'=1000 [\Omega]$.}

\begin{figure}[!h]
\centering
   \subfigure[]{%
     \includegraphics[width=0.98\columnwidth]{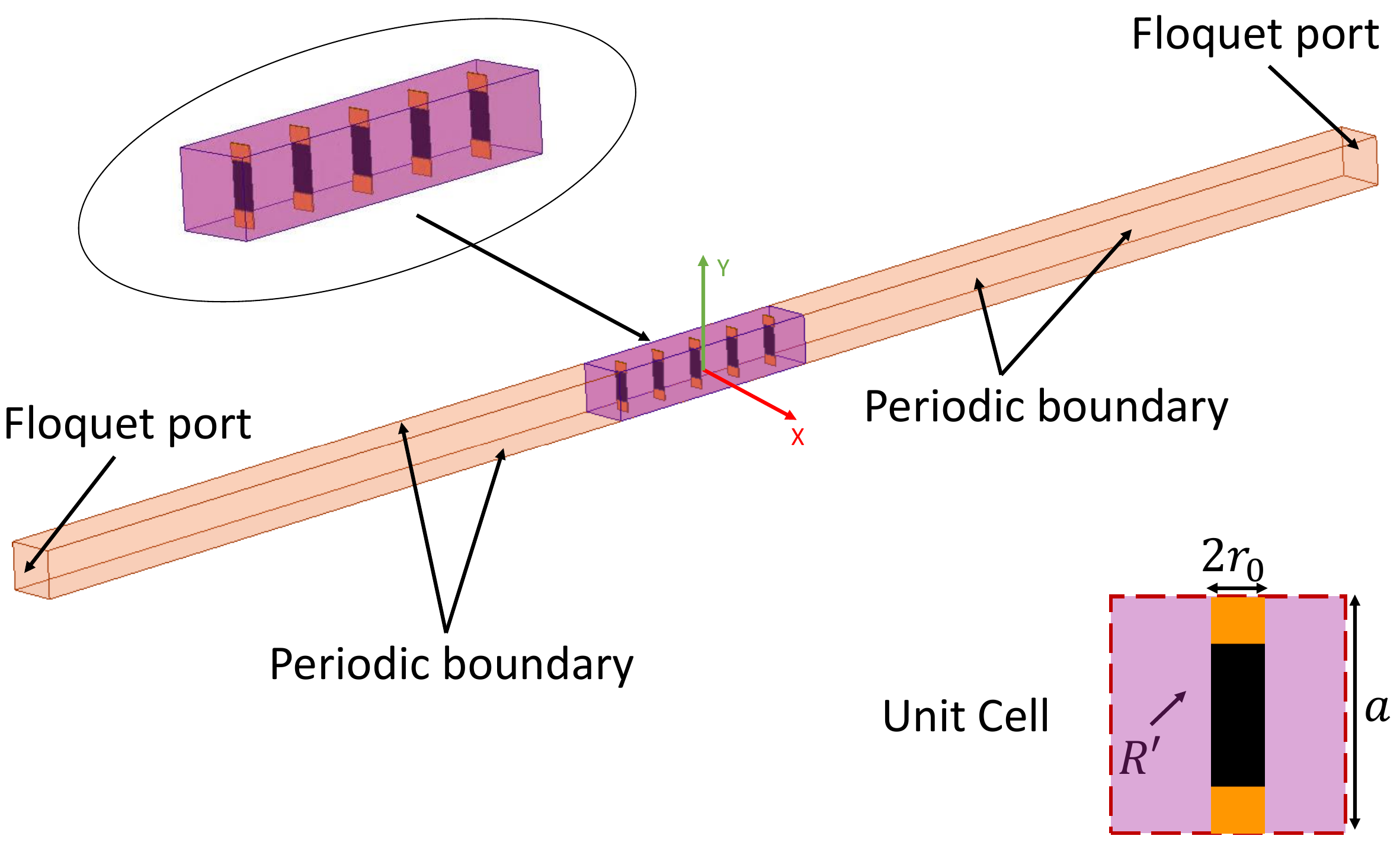}}
\\
   \subfigure[]{%
     \includegraphics[width=0.58\columnwidth]{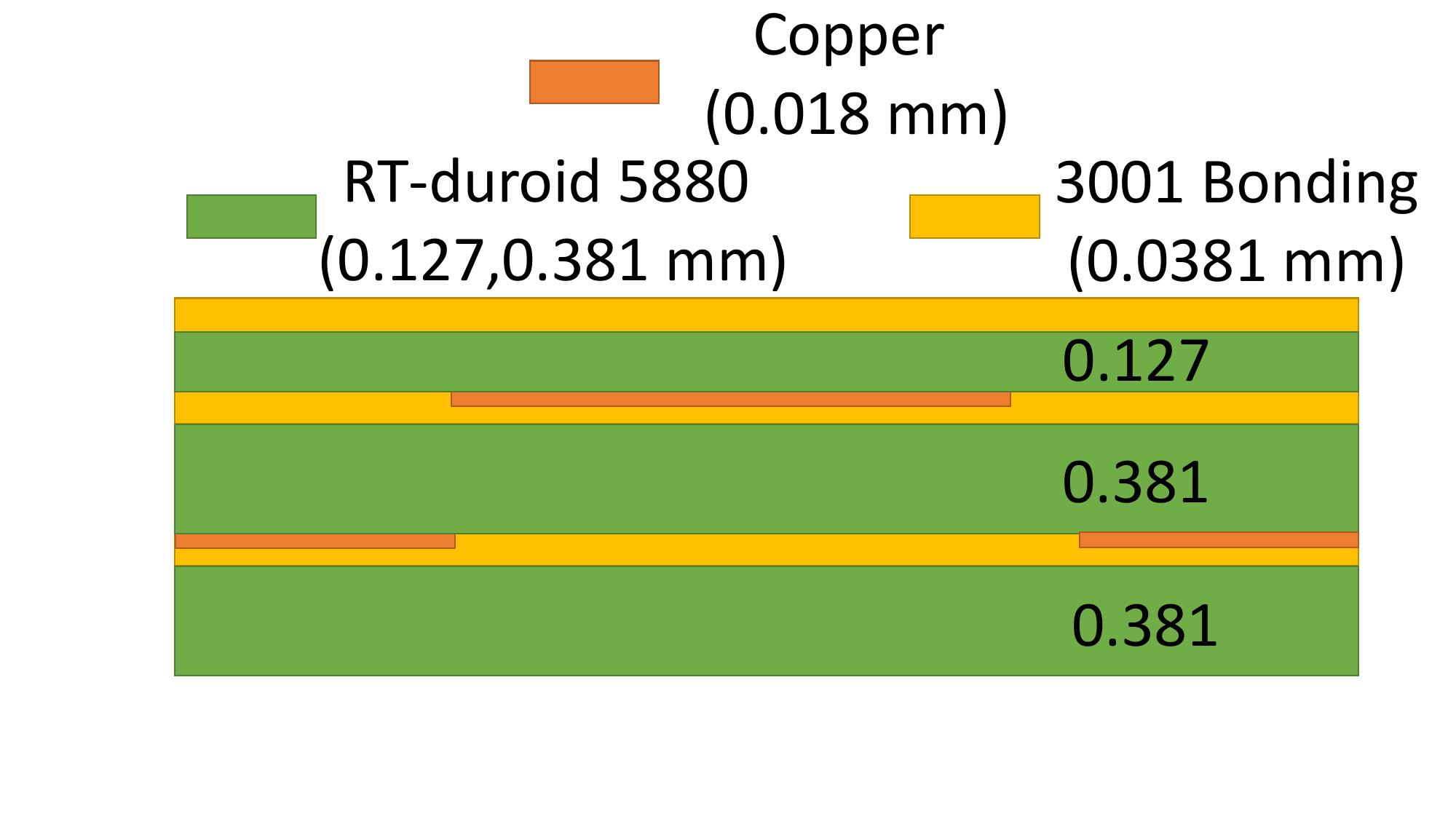}}
   \subfigure[]{%
     \includegraphics[width=0.40\columnwidth]{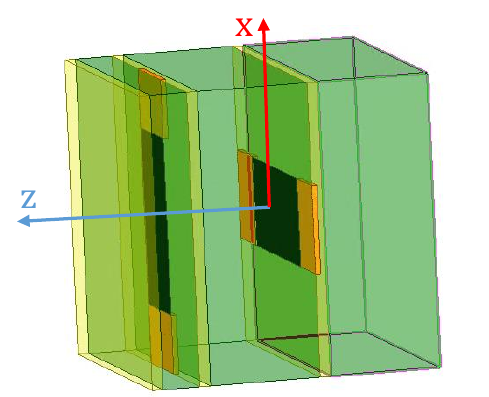}}
\\
    \subfigure[]{%
     \includegraphics[width=0.98\columnwidth]{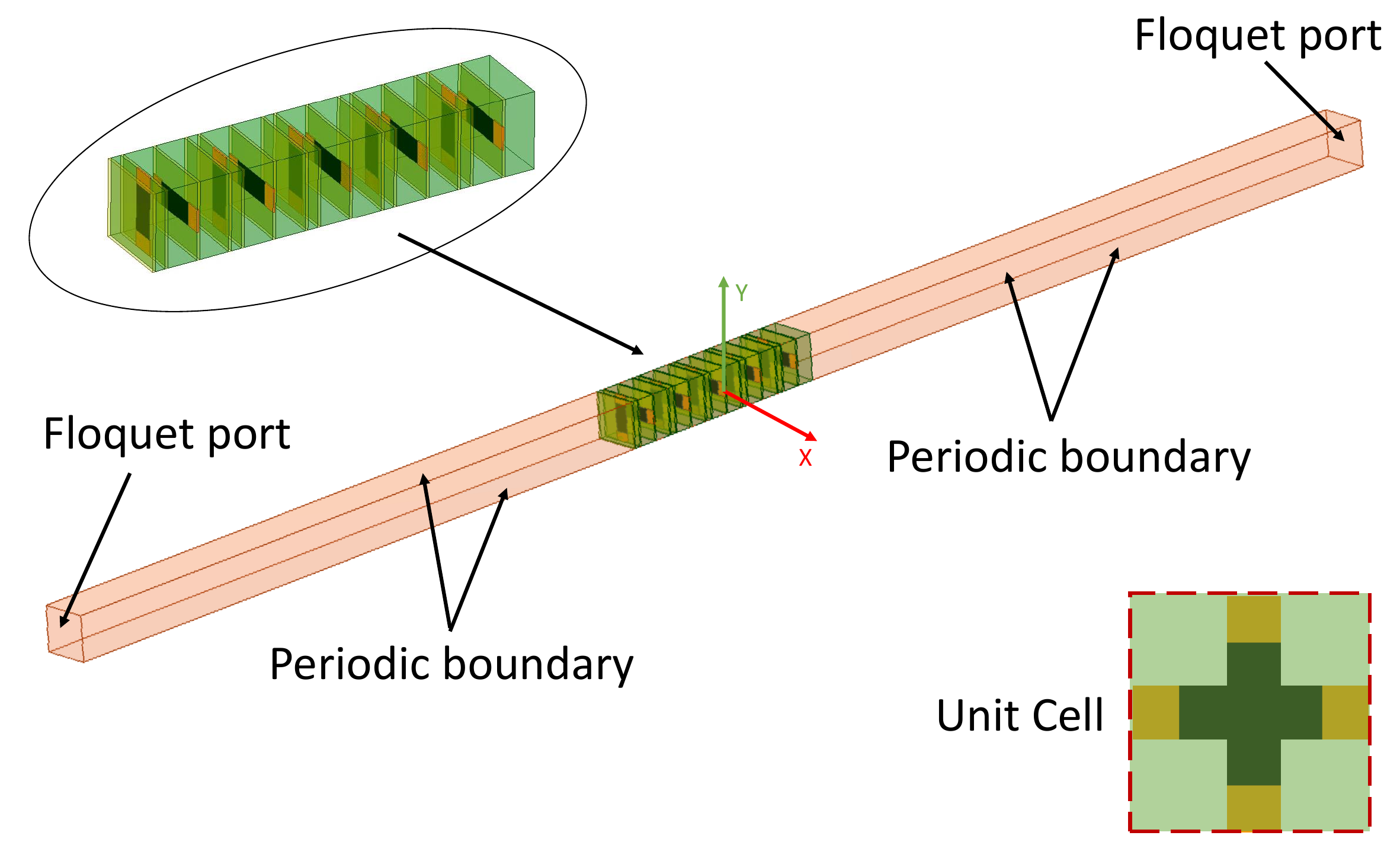}}
\caption{Design and simulation layout in HFSS. (a) The initial simulation, used to extract the necessary host material properties, featured two Floquet ports defined at both ends of the structure. Periodic boundary conditions were enforced on the other four faces of the structure. (b) {Side view of} the stackup of a single unit cell, based on ROGERS  {RT-duroid 5880} lamination and {ROGERS 3001} bonding material. {(c) Oblique view of the unit cell.} (d) The simulation setup of the final structure (similar to Fig.\ref{Metamaterial_Realization} (a)), incorporating the stackup described in (b).
A parametric sweep was conducted on $R'$ to determine the optimal parameter for the design.}
\label{Metamaterial_Realization}
\end{figure}

Step 2 - Realizing an infinite absorbing metasurface structure in HFSS using strips instead of wires for practical purposes. The structure (without the PEC backplane) comprises of 5 unit cells (shown to be a sufficient number of unit cells \cite{Smith2002Determination}), with a total thickness of {$5a=d=0.005[\rm{m}]$} along the Z direction. Periodic boundary conditions were implemented on four faces of the structure, specifically at the XY plane. Additionally, two Floquet ports were assigned to the remaining faces, positioned at a distance of  {$60.5[\rm{mm}]$} from the layer. The structure is excited by the fundamental Floquet wave-field, where the electric field aligns with the strip's direction and propagates normally to the layer, see Fig.~\ref{Metamaterial_Realization} (a) for illustration.  {In this step, the strips are located at the center of the unit cell.} Note that the distance between the Floquet ports and the absorbing structure was reduced for the purpose of providing a clearer visualization in Fig.~\ref{Metamaterial_Realization} (a). Furthermore {a de-embedding process that shifts and calibrates the numerical reference plane of the simulation (the ``Flouquet ports'' in Fig.\ref{Metamaterial_Realization} (a),(d)) into the interface plane between the vacuum and the absorber was applied.} At this stage, we conducted a sweep on $\epsilon_h$ and compared the scattering matrix, specifically $S_{11}$ and $S_{21}$ {following the re-calibration}, with those obtained when substituting the structure with a bulk material having the required parameters $\epsilon_r$ and $\sigma$. This sweeping process revealed that the optimal host material value should be tuned to  {$\epsilon_h=2.3$.} Note that the fundamental requirement for homogenization, ensuring that the unit cell dimension $a$ is significantly smaller than the minimal operating wavelength within the material, given by {$\lambda_m=0.02/\sqrt{2.3}=13.1876 [\rm{mm}]$} is satisfied.

Step 3 - Search for appropriate raw and bonding PCB materials. ROGERS offer  {RT-duroid 5880 laminate material with $\epsilon_d=2.2,\rm{tan}\delta_d=0.0009$ and ROGERS 3001 bonding material with $\epsilon_b=2.28,\rm{tan}\delta_b=0.003$ \cite{ROGERS}.}
The stackup for the structure is established by combining these two materials with copper (simulated as PEC) traces. Figure~\ref{Metamaterial_Realization} (b) illustrates the stackup for a single unit cell {(side view)}.  {Figure~\ref{Metamaterial_Realization} (c) depicts an oblique view of the unit cell structure.}
{During the manufacturing process the bonding material covers the thin metallic strips.} Consequently, the thickness of each unit cell is {$1.0033[\rm{mm}]$}, resulting in a total absorber thickness of {$5.0165 [\rm{mm}]$.}
Fig.~\ref{Metamaterial_Realization} (d) illustrates the final structure, incorporating the simulation definitions outlined in Fig.~\ref{Metamaterial_Realization} (a) and the absorber layout depicted in Figs.~\ref{Metamaterial_Realization} (b) {and (c)}. Unlike Fig.~\ref{Metamaterial_Realization} (a), both horizontal and vertical loaded wires are introduced to ensure polarization insensitivity, specifically for electric fields within the wire plane (TE polarization). Importantly, the wires are positioned on separate planes, with a distance of  {$0.381 [\rm{mm}]$} , as indicated in the stackup.
An additional round of parametric sweep is carried out on the absorbing structure, varying the resistance $R'$ in a manner similar to the preceding step. The electric field polarization was maintained as in the previous sweep. The {corrected} optimal observed value is  {$R'=1025 [\rm{\Omega}]$} .

Step 4 - Homogenization of the final metamaterial design using scattering data extracted by full wave simulations.  The metamaterial shown in Fig.~\ref{Metamaterial_Realization} (c) undergoes homogenization, where its effective parameters are derived from the simulated scattering matrix.
The methodology for this process was initially introduced for normal incidence in \cite{Smith2002Determination}. {Here, we aim to extract the effective parameters not only for normal incidences, but for TE and TM oblique incidences. Therefore, we take inspiration from existing works \cite{Menzel2008Retrieving,Cohen2015Bi} and extract the effective wave parameters under oblique incidence (see Appendix~\ref{AppendixA} for additional information). An alternative approach, though not implemented here, involves directly extracting the permittivity and permeability tensor coefficients using an extended field-sampling method \cite{Watanabe2008S-pol}.}
It is important to note that the technique employed to extract the effective parameters in this work requires knowledge of the scattering matrix at the interface between the metamaterial and its surroundings, {therefore the de-embedding process}. Given the small size of the sample in comparison to the operating wavelength, there is no branching problem \cite{solt2010Unique}.
{The proposed metamaterial shown in Fig. \ref{Metamaterial_Realization}(c) lacks symmetry in the X-Y plane. Consequently, scanning solely over $\theta$ is inadequate for accurately extracting its effective parameters; a scan across the X-Y plane is also required. This critical insight is explored and addressed further in the following discussion.}

\begin{figure}[!h]
\centering
   \subfigure[]{%
     \includegraphics[width=0.49\columnwidth]{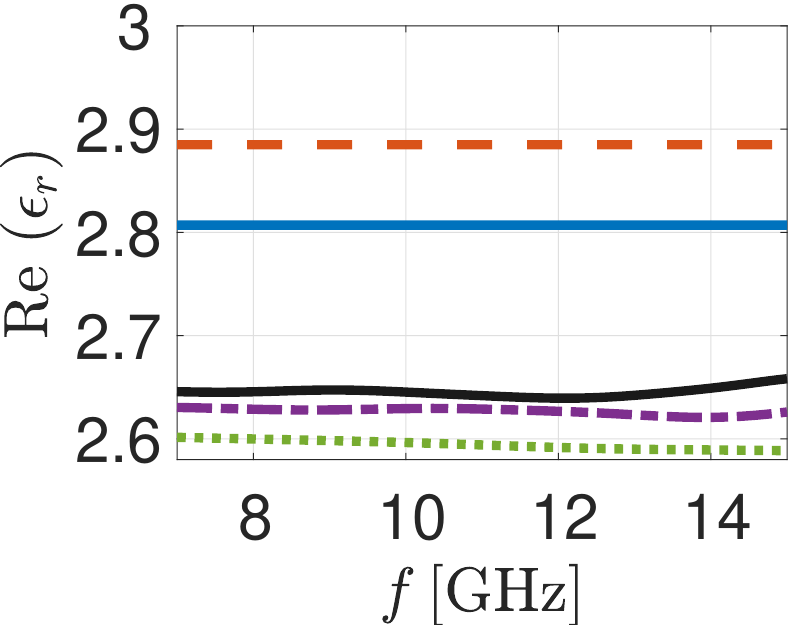}}
    \subfigure[]{%
     \includegraphics[width=0.49\columnwidth]{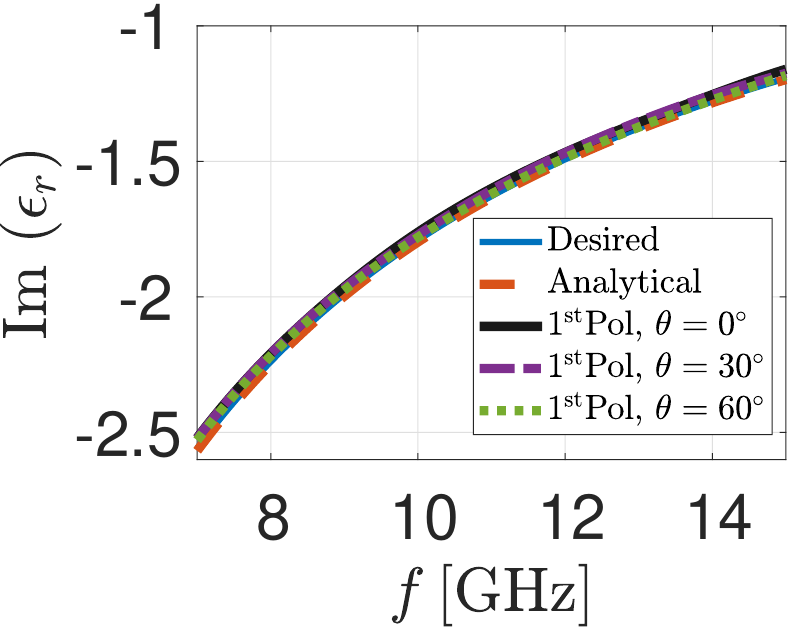}}
\\
   \subfigure[]{%
     \includegraphics[width=0.49\columnwidth]{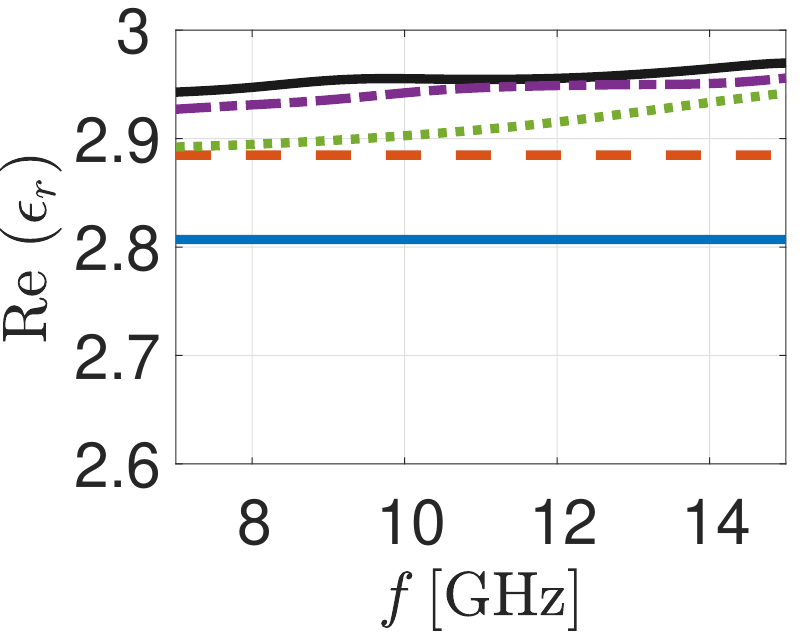}}
   \subfigure[]{%
     \includegraphics[width=0.49\columnwidth]{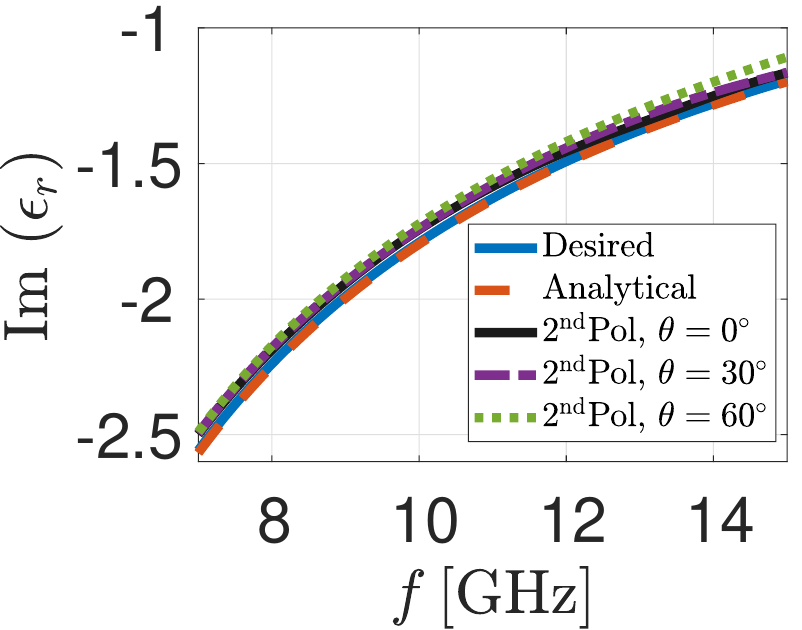}}
\\
    \subfigure[]{%
     \includegraphics[width=0.49\columnwidth]{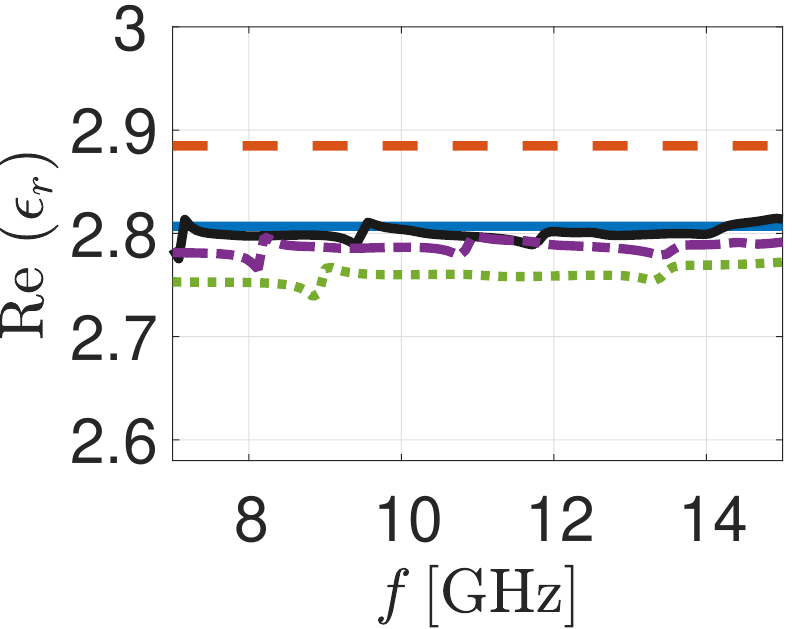}}
    \subfigure[]{%
      \includegraphics[width=0.49\columnwidth]{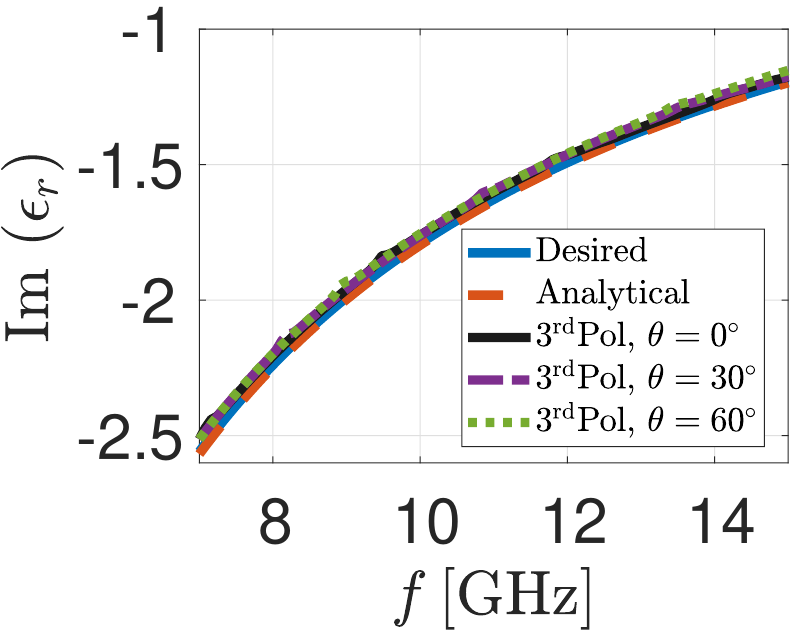}}
\caption{{Comparison between the desired, analytical and extracted permittivities (TE polarization). The desired (blue) and analytical (red) are irrespective of the incidence wave's polarization. The extracted results for $\theta=0^\circ$ (black), $\theta=30^\circ$ (dashed purple) and $\theta=60^\circ$ (dotted green) are presented. (a),(b) Real and imaginary parts under the $1^{\rm{st}}$ Pol. (c),(d) Similar to (a),(b) however for the $2^{\rm{nd}}$ Pol. (e), (f) Similar to (a),(b) however for the $3^{\rm{rd}}$ Pol.}}
\label{Homogenization}
\end{figure}
\begin{figure}[!h]
\centering
   \subfigure[]{%
     \includegraphics[width=0.49\columnwidth]{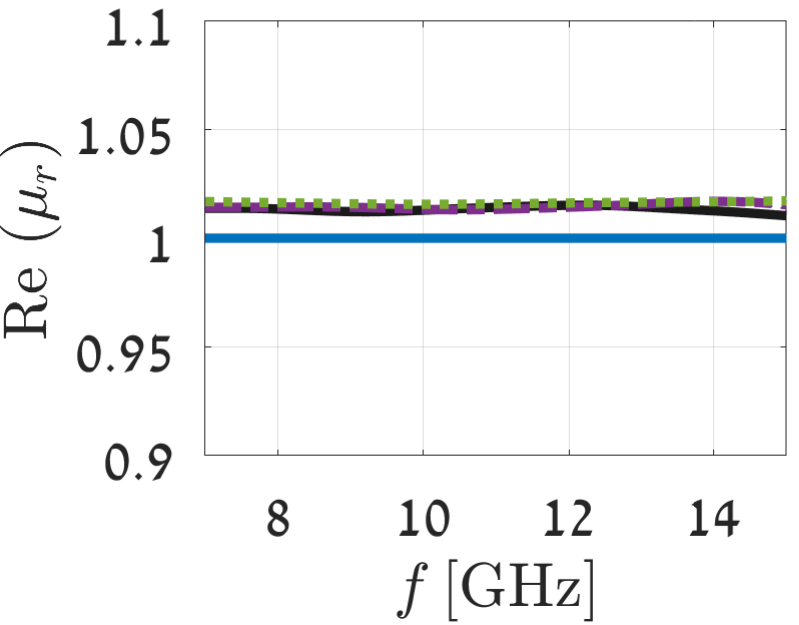}}
    \subfigure[]{%
     \includegraphics[width=0.49\columnwidth]{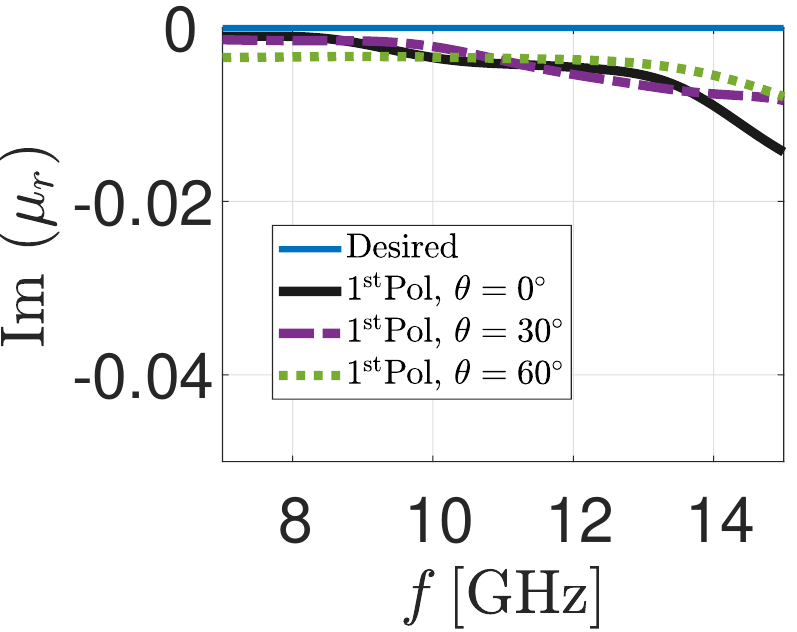}}
\\
   \subfigure[]{%
     \includegraphics[width=0.49\columnwidth]{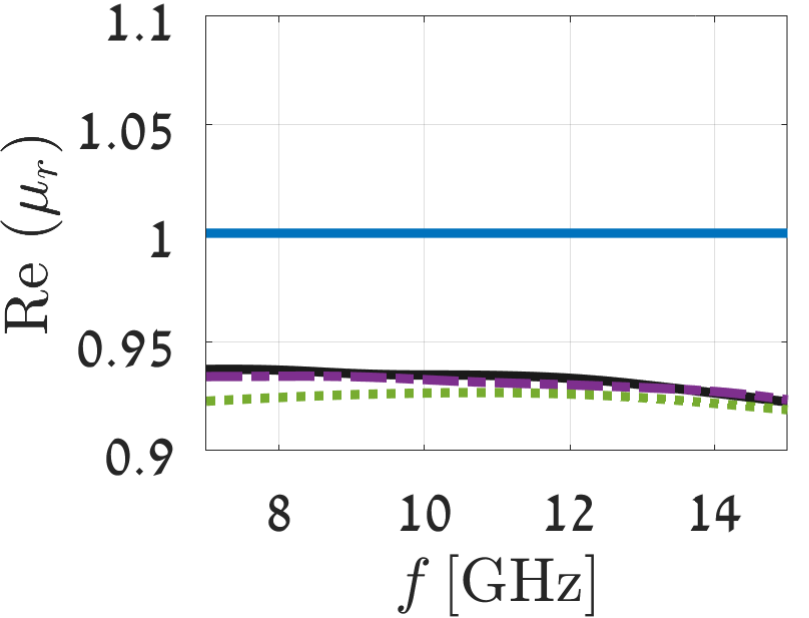}}
   \subfigure[]{%
     \includegraphics[width=0.49\columnwidth]{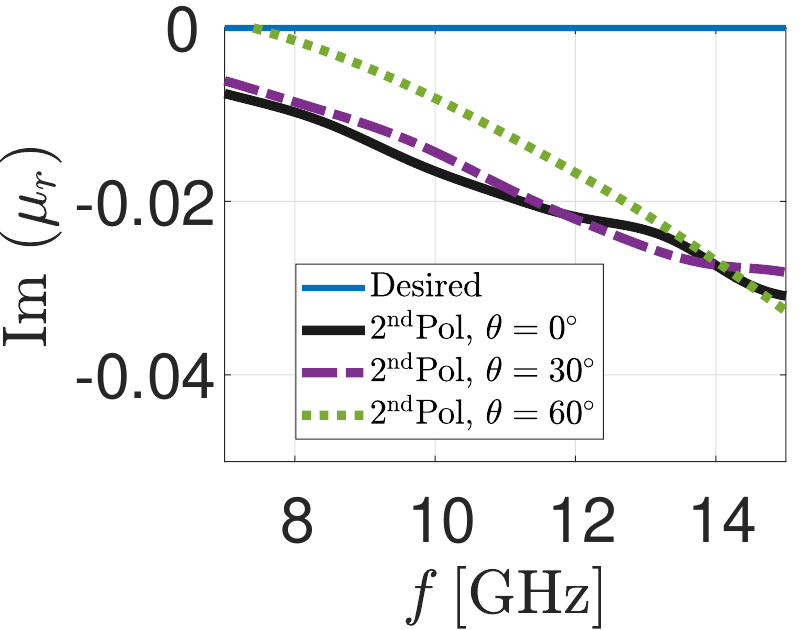}}
\\
    \subfigure[]{%
     \includegraphics[width=0.49\columnwidth]{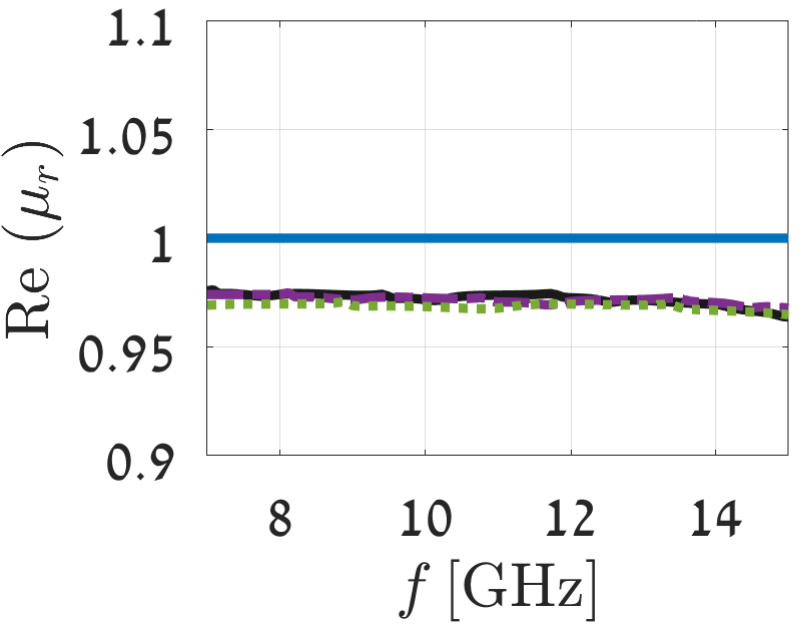}}
    \subfigure[]{%
      \includegraphics[width=0.49\columnwidth]{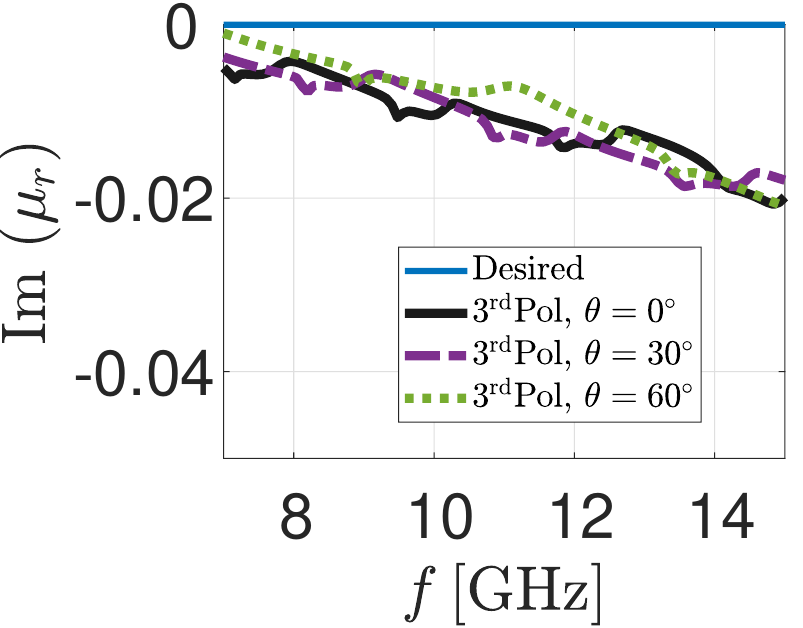}}
\caption{{Comparison between the desired and extracted permeabilities (TE polarization). The desired (blue) and analytical (red) are irrespective of the incidence wave's polarization. The extracted results for $\theta=0^\circ$ (black), $\theta=30^\circ$ (dashed purple) and $\theta=60^\circ$ (dotted green) are presented. (a),(b) Real and imaginary parts under the $1^{\rm{st}}$ Pol. (c),(d) Similar to (a),(b) however for the $2^{\rm{nd}}$ Pol. (e), (f) Similar to (a),(b) however for the $3^{\rm{rd}}$ Pol.}}
\label{Homogenization_mu}
\end{figure}

Our {first} objective is to identify the essential parameters for TE polarization, specifically when the electric field is confined to the XY plane (the plane of the wires). While the general electric field may consist of three components, i.e., $\overrightarrow{E}=E_{x}\hat{x}+E_{y}\hat{y}+E_{z}\hat{z}$, TE polarization satisfy to the following conditions:
\begin{equation}
\label{TE_Cond}
\overrightarrow{E}\cdot\hat{n}=0,\hspace{0.5cm} \overrightarrow{E}\cdot\hat{k}=0.
\end{equation}
Here, $\hat{n}=\hat{z}$ represents the normal to the layer, and $\hat{k}=\rm{sin(\theta)cos(\phi)}\hat{x}+\rm{sin(\theta)sin(\phi)}\hat{y}+\rm{cos(\theta)}\hat{z}$ denotes the direction of propagation for the incident wave, {and $\phi$ denotes the azimuthal angle in the X-Y plane, while $\theta$ denotes an elevation angle measured with respect to the normal}.
Note that $\hat{n}$ and $\hat{k}$ set the plane of incidence. The first condition implies that $E_{z}=0$, confirming the electric field's polarization within the plane of the wires. The subsequent condition implies that $E_{x}\rm{cos(\phi)}+E_{y}\rm{sin(\phi)}=0$ (for $\theta \neq 0$), which sets the relation between $E_{x}$ and $E_{y}$.
We define three distinct cases to extract the effective parameters: In the first case, when {$\phi=0$}, the electric field is $\hat{y}$ polarized, denoted as $1^{\rm{st}}$ Pol. In the second case, for {$\phi=\pi/2$}, the electric field becomes $\hat{x}$ polarized, identified as $2^{\rm{nd}}$ Pol. Lastly, in the third case with $\phi=\pi/4$, the electric field exhibits diagonal polarization at $45^\circ$ from the $x$ and $y$ axes, denoted as $3^{\rm{rd}}$ Pol.
Figure~\ref{Homogenization} presents the homogenization results {for TE polarization}, with the desired (see Eq.\eqref{eps_desired}) and analytical (see Eq.\eqref{eps_h}) permittivities represented by blue and red dashed lines, respectively, {which are desired to be independent} of the incidence wave's polarization. Subfigures (a) and (b) illustrate the real and imaginary parts of the extracted relative permittivity for the $1^{\rm{st}}$ Pol at $\theta=0^\circ$ (black), $\theta=30^\circ$ (dash-dotted purple), and $\theta=60^\circ$ (dotted green). Subfigures (c) and (d) depict the corresponding results for the $2^{\rm{nd}}$ Pol, while subfigures (e) and (f) show the results for the $3^{\rm{rd}}$ Pol.
{As shown in Figure~\ref{Homogenization}, the designed metamaterial achieves relative (real) permittivities in the range  $\epsilon_r \in[2.6,3]$ across the entire frequency range of operation, while its imaginary part closely matches the desired profile. To provide the complete picture, Fig.~\ref{Homogenization_mu} depicts the extracted permeability $\mu_r$ for the three excitations, compared to the desired value of $\mu_r=1$. The real part of the permeability ranges from $0.919$ to $1.02$, while the imaginary part remains small.  }

\begin{figure}[!h]
\centering
   \subfigure[]{%
     \includegraphics[width=0.49\columnwidth]{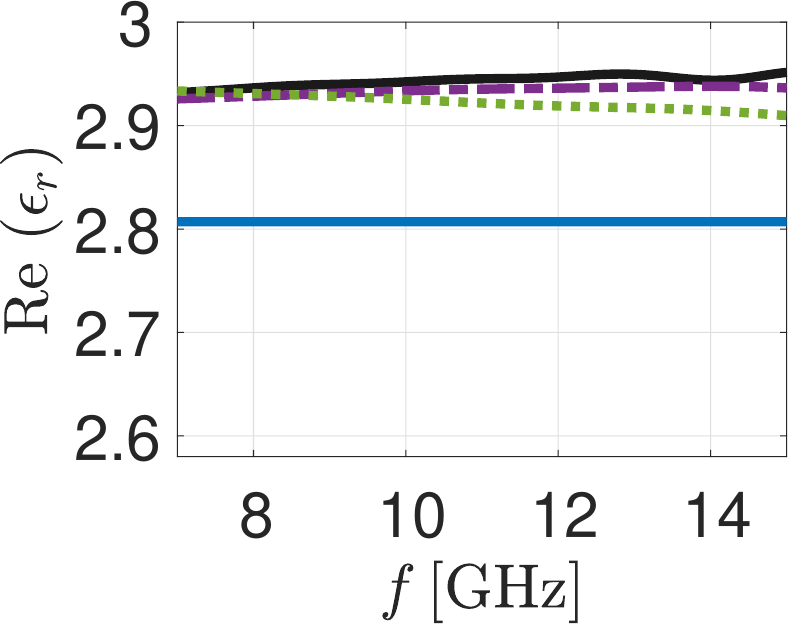}}
    \subfigure[]{%
     \includegraphics[width=0.49\columnwidth]{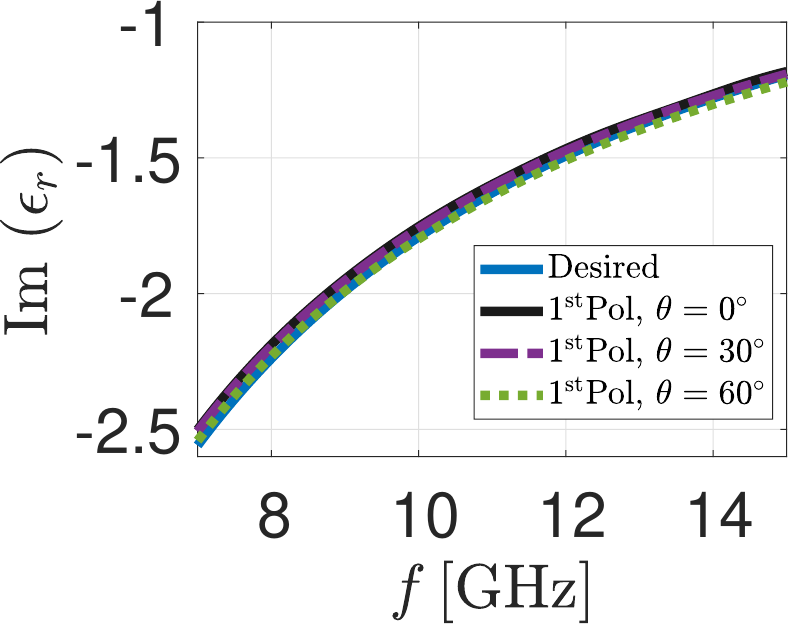}}
\\
   \subfigure[]{%
     \includegraphics[width=0.49\columnwidth]{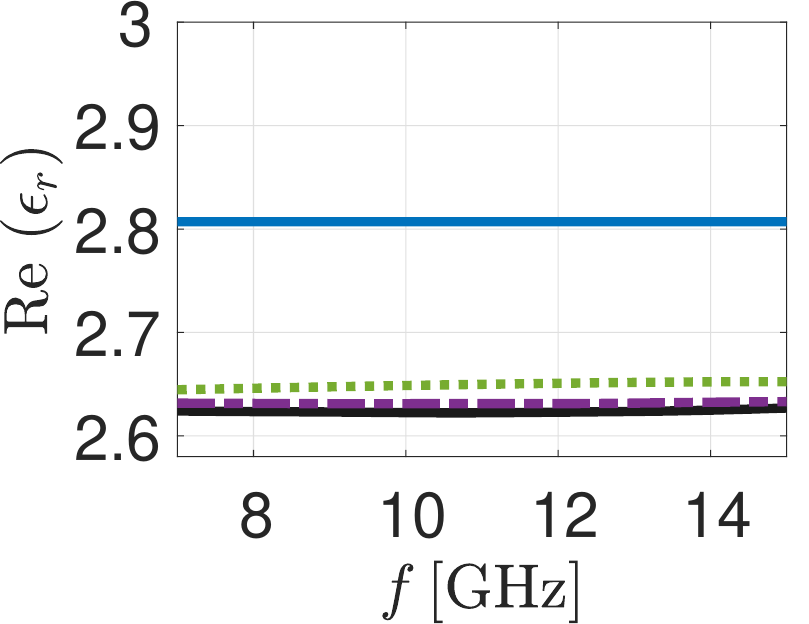}}
   \subfigure[]{%
     \includegraphics[width=0.49\columnwidth]{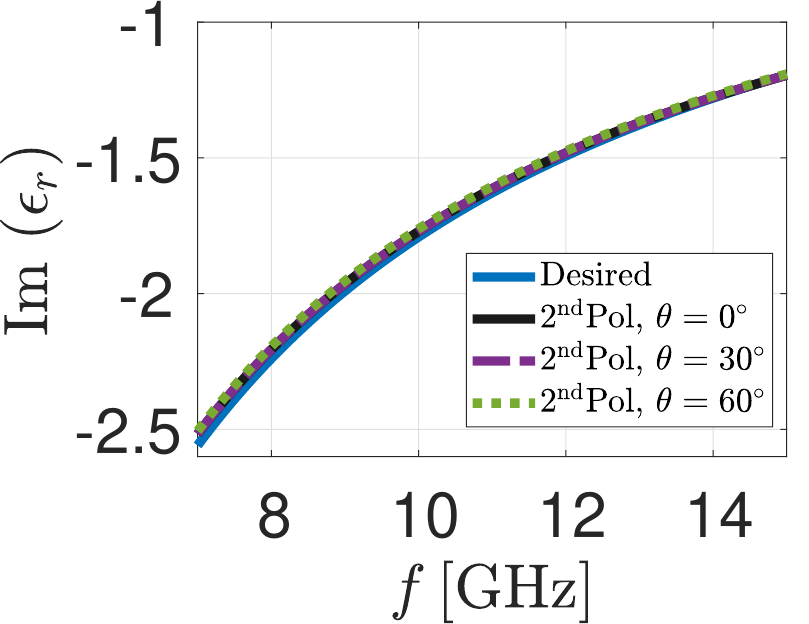}}
\\
    \subfigure[]{%
     \includegraphics[width=0.49\columnwidth]{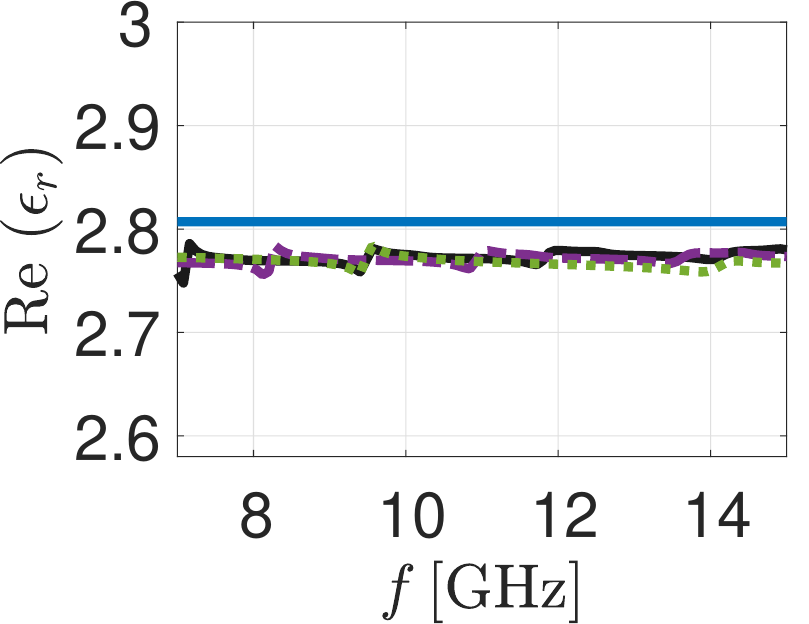}}
    \subfigure[]{%
      \includegraphics[width=0.49\columnwidth]{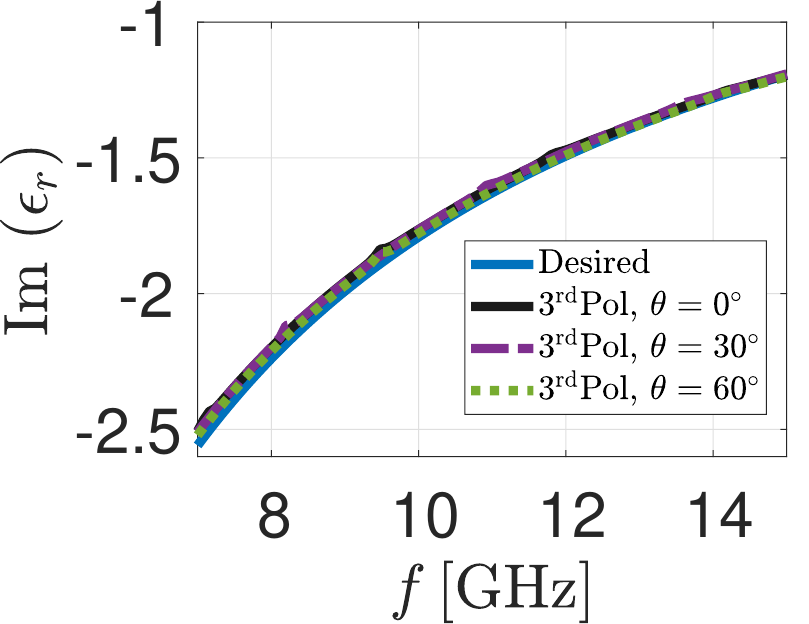}}
\caption{{Comparison between the desired, analytical and extracted permittivities (TM polarization). The desired (blue) is irrespective of the incidence wave's polarization. The extracted results for $\theta=0^\circ$ (black), $\theta=30^\circ$ (dashed purple) and $\theta=60^\circ$ (dotted green) are presented. (a),(b) Real and imaginary parts under the $1^{\rm{st}}$ Pol. (c),(d) Similar to (a),(b) however for the $2^{\rm{nd}}$ Pol. (e), (f) Similar to (a),(b) however for the $3^{\rm{rd}}$ Pol.}}
\label{Homogenization_TM_eps}
\end{figure}

\begin{figure}[!h]
\centering
   \subfigure[]{%
     \includegraphics[width=0.49\columnwidth]{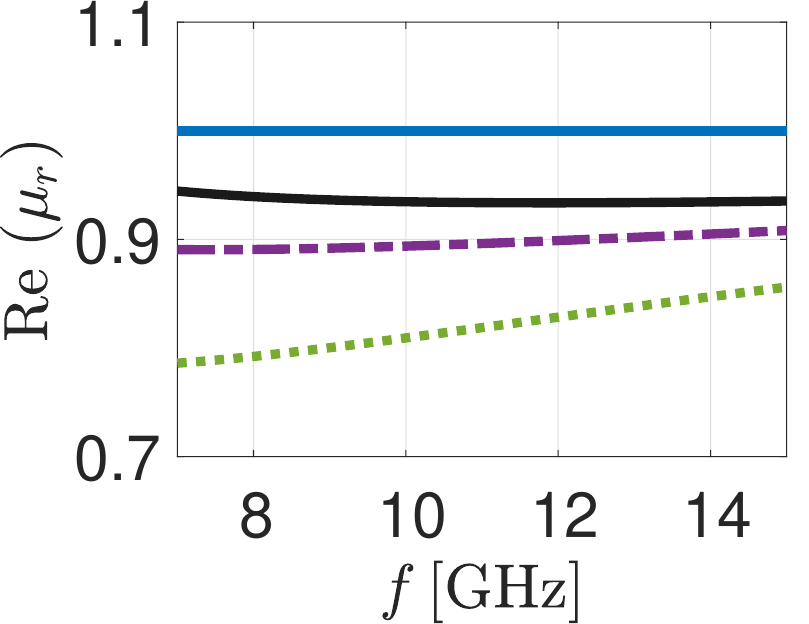}}
    \subfigure[]{%
     \includegraphics[width=0.49\columnwidth]{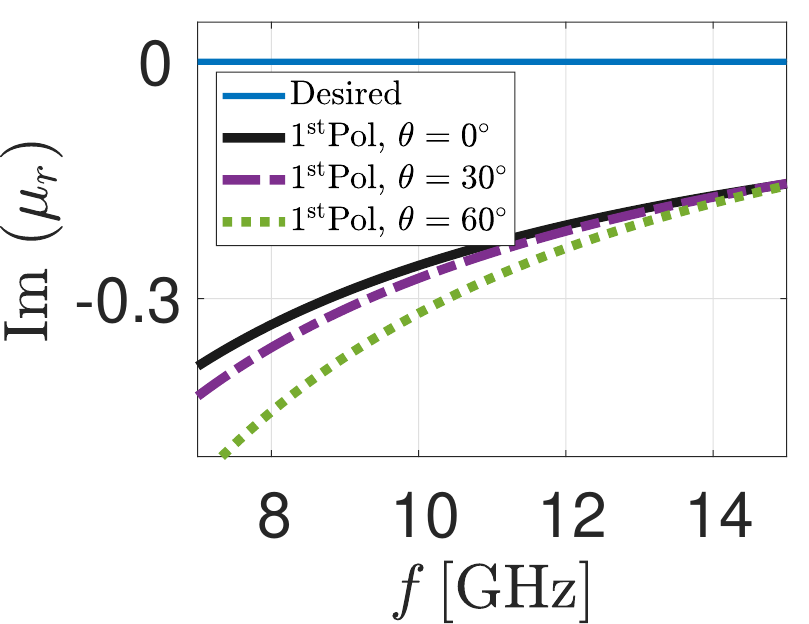}}
\\
   \subfigure[]{%
     \includegraphics[width=0.49\columnwidth]{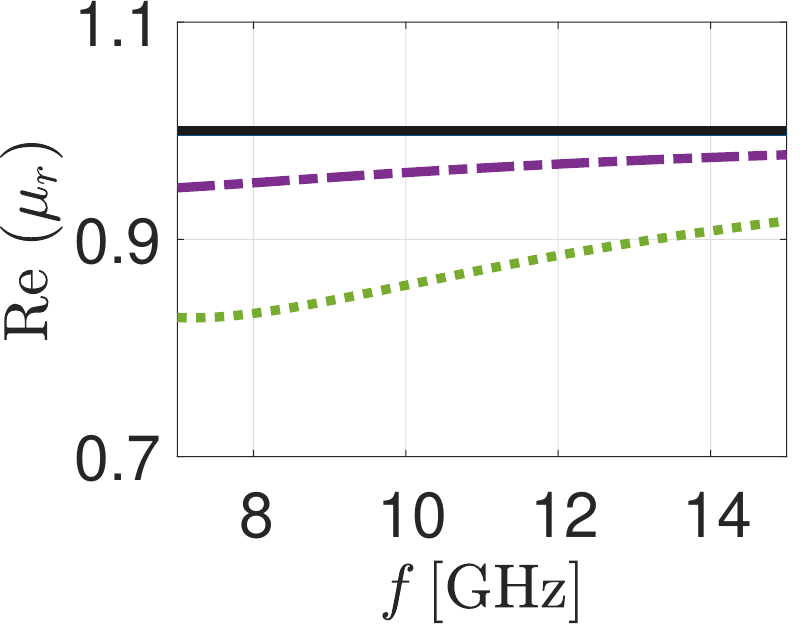}}
   \subfigure[]{%
     \includegraphics[width=0.49\columnwidth]{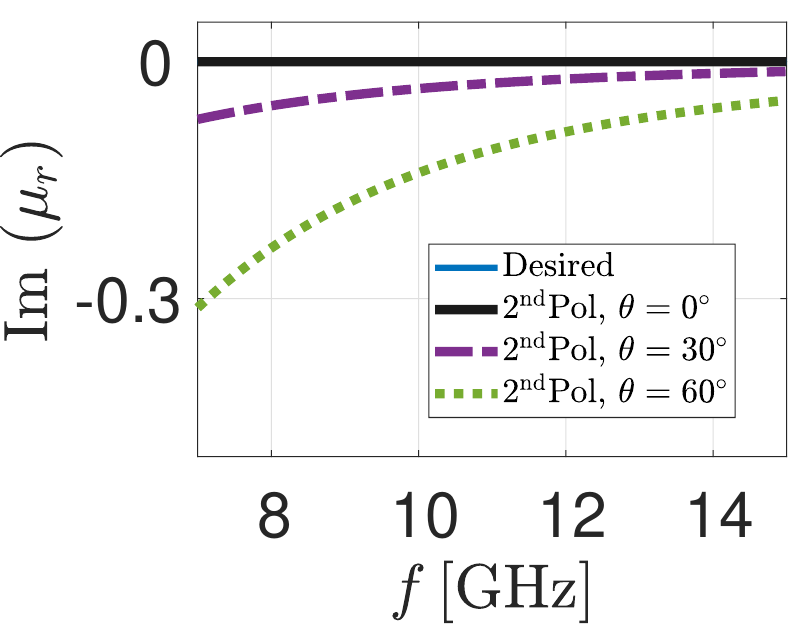}}
\\
    \subfigure[]{%
     \includegraphics[width=0.49\columnwidth]{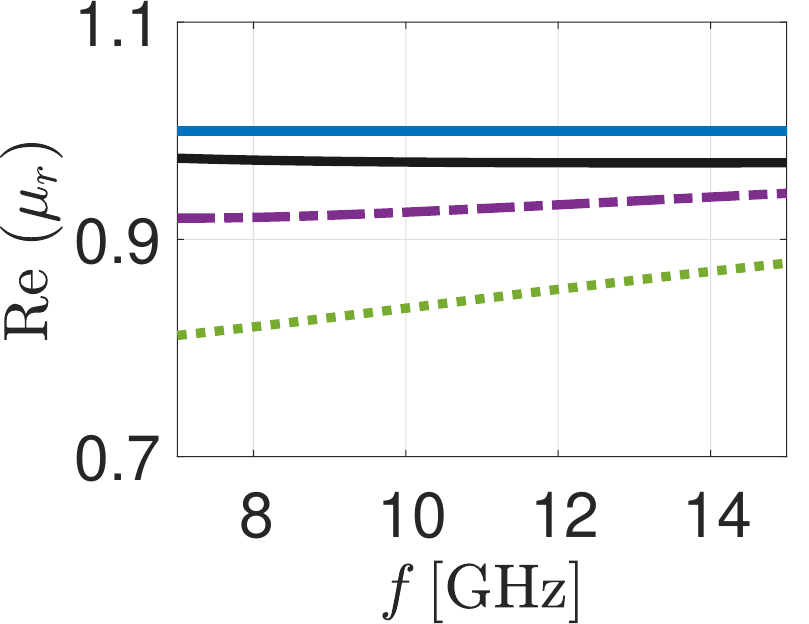}}
    \subfigure[]{%
      \includegraphics[width=0.49\columnwidth]{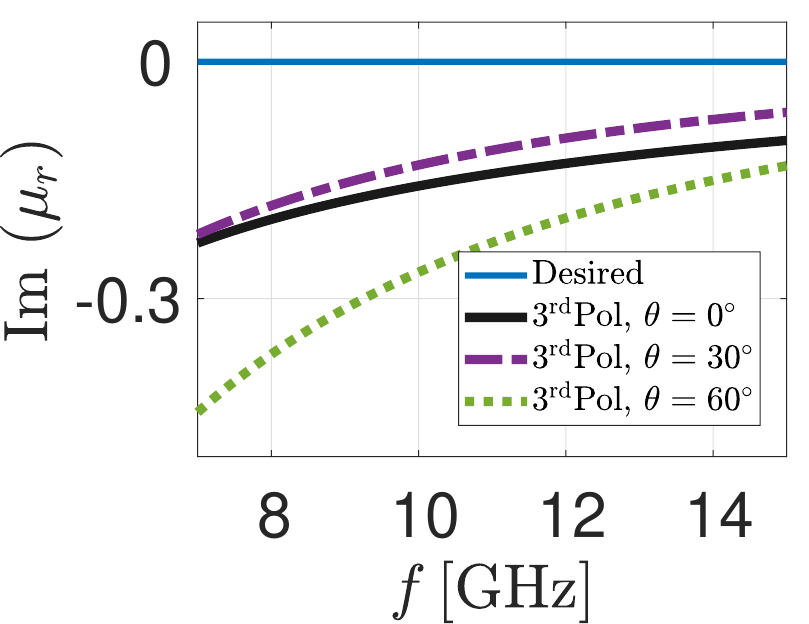}}
\caption{{Comparison between the desired and extracted permeabilities (TM polarization). The desired (blue) is irrespective of the incidence wave's polarization. The extracted results for $\theta=0^\circ$ (black), $\theta=30^\circ$ (dashed purple) and $\theta=60^\circ$ (dotted green) are presented. (a),(b) Real and imaginary parts under the $1^{\rm{st}}$ Pol. (c),(d) Similar to (a),(b) however for the $2^{\rm{nd}}$ Pol. (e), (f) Similar to (a),(b) however for the $3^{\rm{rd}}$ Pol.}}
\label{Homogenization_mu_TM}
\end{figure}

{Our next objective is to extract the parameters of the metamaterial structure under TM illumination.
Specifically, the magnetic field is confined to the XY plane (the plane of the wires). The magnetic field of such polarization satisfies the relations in Eq.~\eqref{TE_Cond} with the replacement $E \rightarrow H$.
The polarization definition for the TM case is defined similarly to the TE case. When $\phi=0$, the magnetic field is $\hat{y}$ polarized, denoted as $1^{\rm{st}}$ Pol. In the second case, for $\phi=\pi/2$, the electric field becomes $\hat{x}$ polarized, identified as $2^{\rm{nd}}$ Pol. Lastly, in the third case with $\phi=\pi/4$, the electric field exhibits diagonal polarization at $45^\circ$ from the $x$ and $y$ axes, denoted as $3^{\rm{rd}}$ Pol. Figures~\ref{Homogenization_TM_eps} and ~\ref{Homogenization_mu_TM} depict the extracted relative permittivities and permeabilities for all three polarizations, respectively. The blue line shows the desired result. The extracted results are presented for three values of $\theta$: a solid black line for $\theta=0^\circ$, a dashed purple line for $\theta=30^\circ$, and a dotted green line for $\theta=60^\circ$.
To ensure that the imaginary part of the permeability corresponds its real part (and also to avoid numerical errors), i.e. satisfies Kramers-Kronig relations, we used a Lorentzian model for the permeability.
The extracted complex permittivity and permeability exhibit good alignment with the desired results. However, at larger angles (e.g., $\theta=60^\circ$), the alignment weakens for the TM case, in contrast to the TE case, where the alignment remains strong even at large angles.}
\subsection{Dallenbach absorber performance}
\begin{figure}[!h]
\centering
   \subfigure[]{%
     \includegraphics[width=0.98\columnwidth]{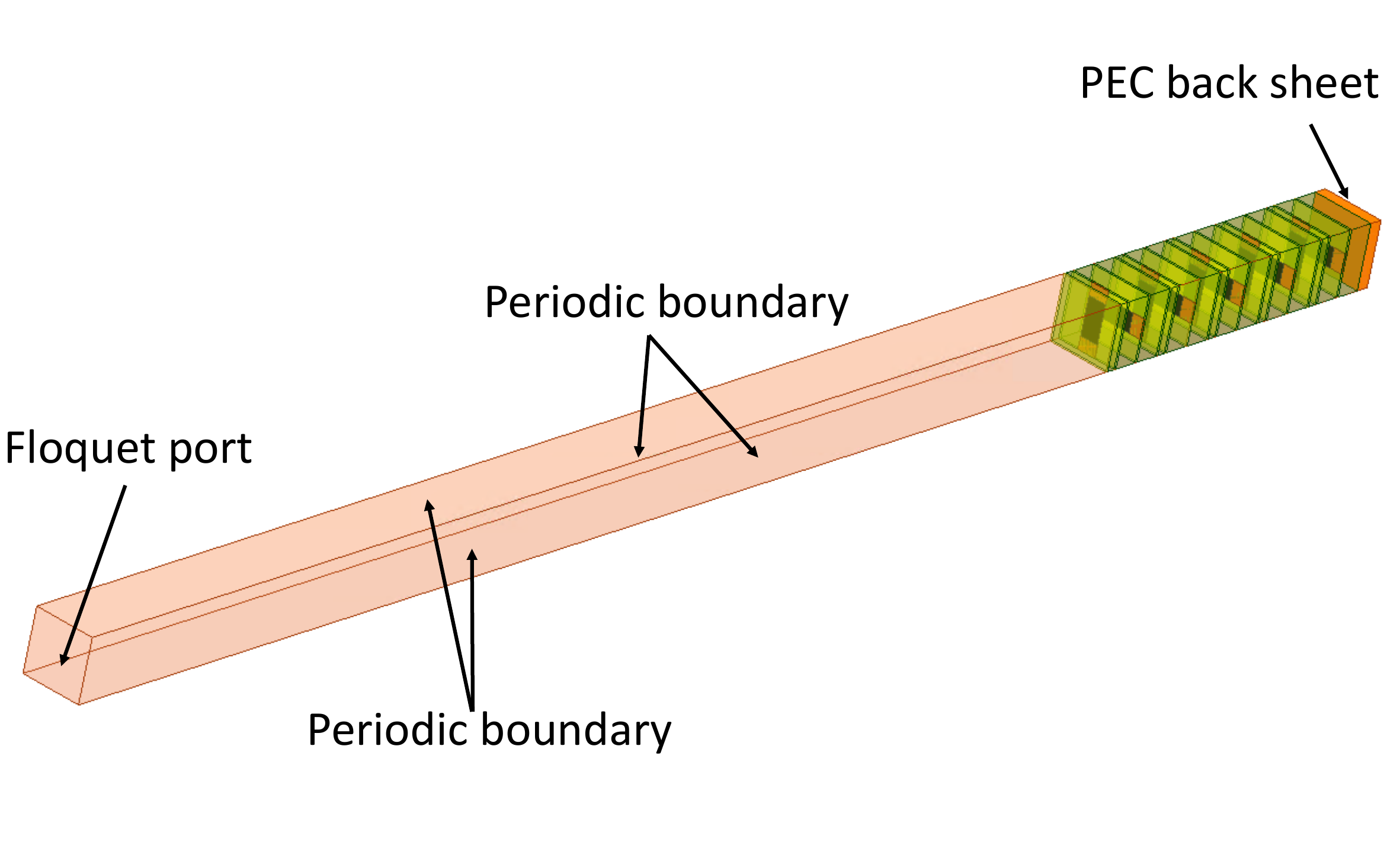}}
\\
   \subfigure[]{%
     \includegraphics[width=0.50\columnwidth]{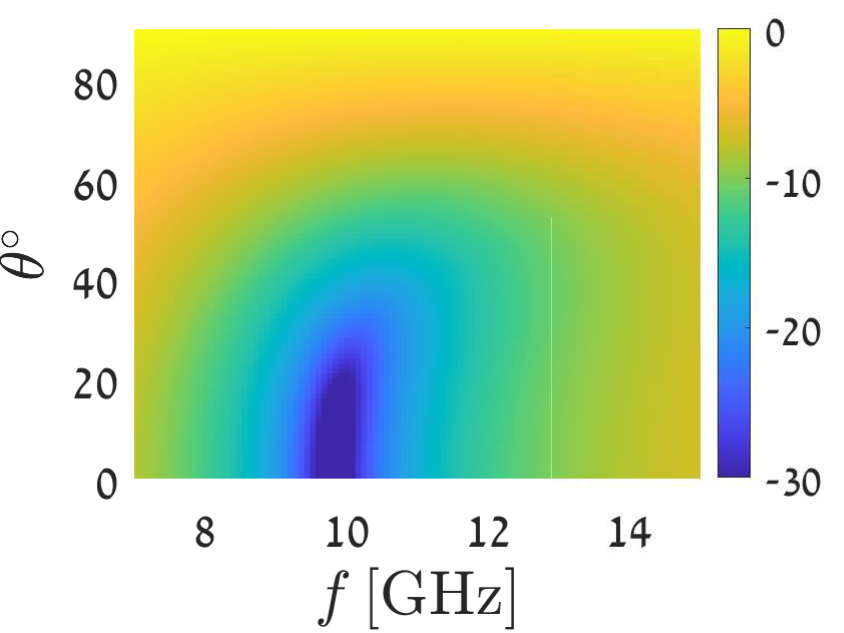}}
    \subfigure[]{%
     \includegraphics[width=0.48\columnwidth]{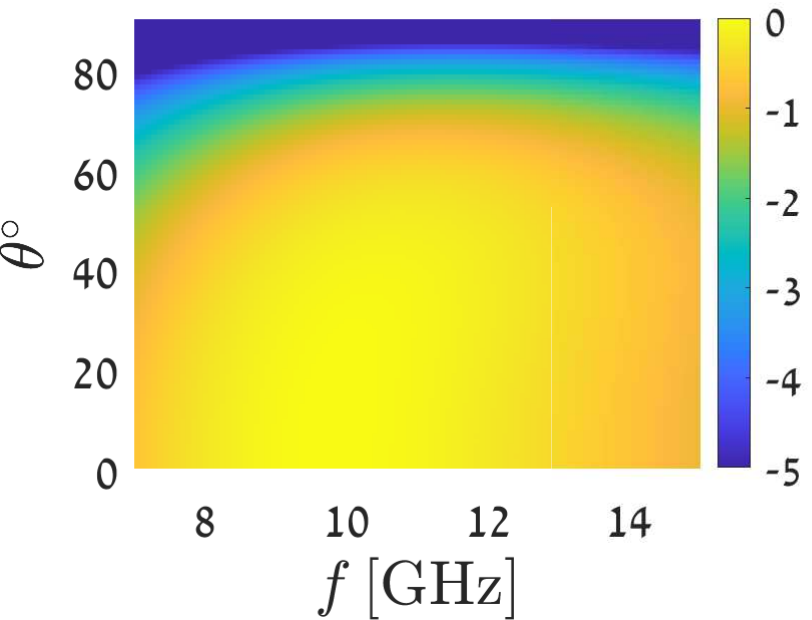}}
\\
   \subfigure[]{%
     \includegraphics[width=0.50\columnwidth]{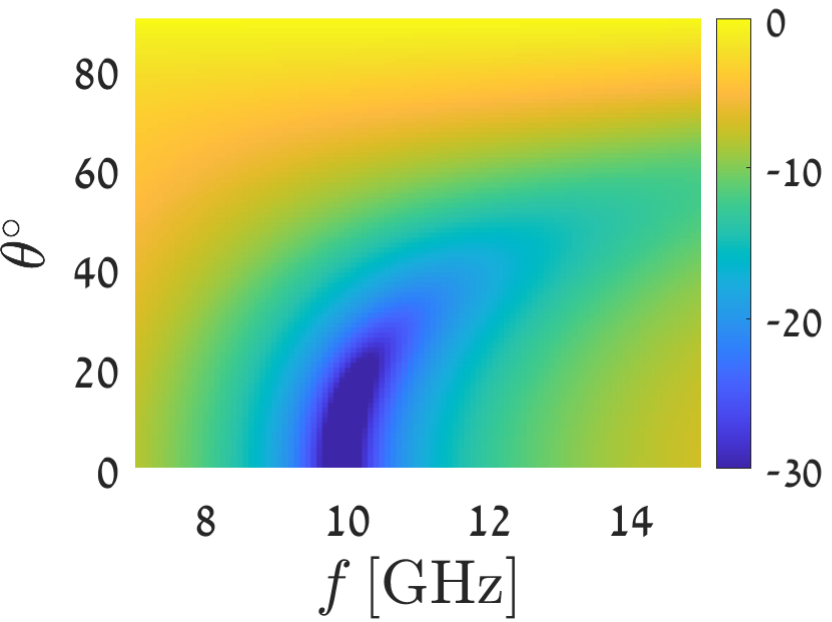}}
    \subfigure[]{%
     \includegraphics[width=0.48\columnwidth]{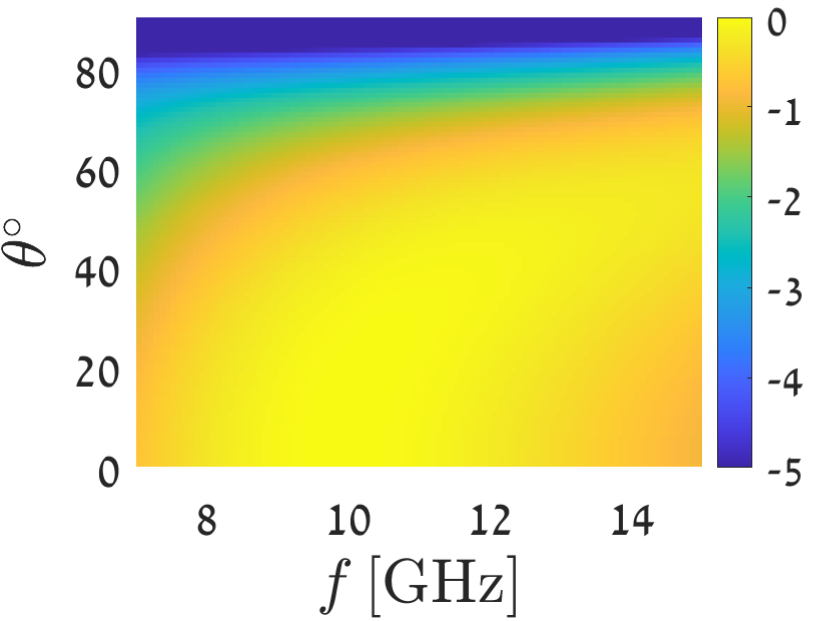}}
\caption{Reflection and absorption characteristics of the Dallenbach metamaterial absorber. (a) Simulation setup in HFSS. (b) TE polarization reflection coefficient, expressed as $20\log_{10} |S_{11}|$, is computed across a range of frequencies between {$7$ and $15$ $\rm{GHz}$} , along with elevation angles spanning from $0$ to $\pi/2$. (c) The absorption coefficient, directly derived from the reflection coefficient, is given by $A=1-|S_{11}|^2$ and presented on a logarithmic scale. {(d) TM reflection map (analogous to (b)). (e) TM absorption map (analogous to (c)).} }
\label{Colormap_abs}
\end{figure}

In the preceding discussion, we introduced a homogenization approach for a metamaterial absorber designed for PCB manufacturing. The objective of this section is to validate the performance of the proposed Dallenbach absorber, assessing both its bandwidth and angular spectrum. To achieve this, we eliminated the back Floquet port from the simulation setup and introduced a PEC back sheet, resulting in a simulation with only one port. The simulation setup is depicted in Fig.~\ref{Colormap_abs} (a).
The reflection coefficient ($S_{11}$) was computed for all three polarizations {(both for TE and TM)}. For each polarization, the reflection coefficient was determined across a grid of 101 equally spaced frequencies ranging from {$7$ to $15$ $\rm{GHz}$} and 91 equally spaced elevation angles spanning from $0$ to $\pi/2$.
Figure~\ref{Colormap_abs} (b) illustrates the reflection {for TE polarized wave}, $20\log_{10} |S_{11}|$, from the designed Dallenbach absorber. The graph presents HFSS simulation results for the $3^{\rm{rd}}$ Pol, with similar {results} observed for other polarizations and when the metamaterial was replaced by a bulk material containing the optimal $\epsilon,\sigma$ obtained in Sec.~\ref{Opt_Approach}. The maximum error between the optimized bulk and its metamaterial realization (for all three polarizations) was {$-24$ db}. {As an additional validation step, we used the effective permittivity and permeability from Figures \ref{Homogenization} and \ref{Homogenization_mu} to create a hypothetical isotropic bulk material. The reflection and absorption coefficients were similar to those shown in Figure \ref{Colormap_abs}, so we have not included them here.} Figure \ref{Colormap_abs} (c) illustrates the absorption coefficient, denoted as $A =1-|S_{11}|^2$, represented on a logarithmic scale ($10 \log_{10} A$).
{Figures \ref{Colormap_abs}(d) and (e) show the corresponding results for the TM case. The maximum error between the optimized bulk material and the proposed metamaterial is approximately $-10 , \text{dB}$, as the metamaterial exhibits increased frequency dependence under TM polarization at larger angles (e.g., $\theta > 50^\circ$). Despite the higher error, the absorption performance remains satisfactory. }
The efficiency of the proposed Dallenbach metamaterial absorber is evident in both frequency and angular spectrums, as depicted in Figure~\ref{Colormap_abs}.

\section{Conclusion}
In this work, we considered the problem of Dallenbach absorbing layers subjected to a wide angular spectrum of oblique incidence plane waves with TE or TM polarizations. A scenario that is typical in near field absorbing applications. We derived a modified sum-rule bound and showed that it is tight in a large parameter range. Moreover, we used this bound derivation in order to formulate an optimization problem for the design of a Dallenbach absorber that is applicable for spectrally wide and broadband impinging wave. A finite PCB realization using effective field theory and backed by full-wave simulations has been proposed. We demonstrated the {efficacy} of the absorber {in three examples} by placing a dipole antenna {operating in the 6G frequency band above a Dallenbach absorber}. {In the first example the dipole antenna was placed {horizontally} more than} a wavelength above {the ground plane}, showing that it can be used to prevent major nulls in the radiation pattern due to destructive interference between direct and reflected waves in the far-field, and in a wide frequency range suitable for {6G} communication applications. {In the second example, we demonstrated that when the dipole antenna is placed in the immediate vicinity of the absorber, the gain variation is reduced bringing the pattern to be more uniform as is desired for wide communication coverage. {In the third example we examined the case of a vertically-oriented dipole antenna placed about a wavelength above the ground plane (similarly to the first case).} In all these scenarios, the designed optimal absorber significantly improved the radiation properties compared to the case where the radiators are placed near a conductor without the proposed optimal absorber. 
These examples demonstrate the applicability of the suggested bound and optimal design approach for timely real-world scenarios.}

\begin{acknowledgments}
C. F.  would like to thank to the Darom Scholarships and High-tech, Bio-tech and Chemo-tech Scholarships, to Yaakov ben-Yitzhak Hacohen excellence Scholarship and to IEEE Antennas and Propagation Society Fellowship (APSF). This research was supported by the Israel Science Foundation (grant No. 1457/23).
\end{acknowledgments}


\appendix
\section{Formulas for extracting effective parameters} \label{AppendixA}
This appendix provides the formulas for parameters extraction under TE and TM polarizations.  The extraction expressions are inspired by, but not identical to \cite{Menzel2008Retrieving}. This is since we use \cite{Cohen2015Bi} to correct a seemingly typographical inconsistency in \cite{Menzel2008Retrieving} that applies for TM polarization. The two port system is described using the S-parameters. The normal wave vector component $k_z$ and the generalized impedances ($\xi_{\rm{TE}}$,$\xi_{\rm{TM}}$) are given by,
\begin{subequations}
\begin{align}
    k_zd\! &=\! \pm \cos^{-1}\!\!\left(\frac{[k_{z}(1\!-\!S_{11}^2)\!+\!k_{z}S_{21}^2]/S_{21}}{k_{z}(1\!-\!S_{11})\!+\!k_{z}(1\!+\!S_{11})}\right)\!\!+\!2\pi m \label{eq:part_a} \\ 
    \xi_{\mathrm{TE}} \! &=\! k_{z}\sqrt{\frac{(1-S_{11})^2-(S_{21})^2}{(1+S_{11})^2-(S_{21})^2}} \label{eq:part_b}   \\ 
    \xi_{\mathrm{TM}} \! &=\! \frac{1}{k_{z}} \sqrt{\frac{(1-S_{11})^2-(S_{21})^2}{(1+S_{11})^2-(S_{21})^2}} \label{eq:part_c}
\end{align}\label{kz_xi}
\end{subequations}
where $m$ is an integer, and is equal to zero for thin layers.  The effective relative parameters are given by,
\begin{subequations}
\begin{align}
    \mu_r &= \frac{k_{z}}{\xi_{\mathrm{TE}}},\hspace{0.1cm} \epsilon_r=\frac{c^2}{\omega^2}\frac{k_t^2+k_z^2}{\mu_r}, \hspace{0.85 cm} \text{TE} \label{eq:TE} \\ 
    \epsilon_r &=k_{z}\xi_{\mathrm{TM}} ,\hspace{0.1cm} \mu_r=\frac{c^2}{\omega^2}\frac{k_t^2+k_z^2}{\epsilon_r}, \hspace{0.5 cm} \text{TM} \label{eq:TM} 
\end{align}
\label{EFF_TETM}
\end{subequations}
where $k_t$ is the transverse component of the wave vector (on the plane normal to $z$).


\begin{thebibliography}{1}
\bibitem{ruck1970radar}
G. Ruck, \emph{Radar Cross Section Handbook}. Springer, 1970.

\bibitem{Sohl2007Physical}
C. Sohl, M. Gustafsson, and G. Kristensson, ``Physical limitations on metamaterials: restrictions on scattering and absorption over a frequency interval,'' Journal of Physics D: Applied Physics, vol. 40, no. 22, pp. 7146–7151, 2007.

\bibitem{Landy2008Perfect}
N. I. Landy, S. Sajuyigbe, J. J. Mock, D. R. Smith, and W. J. Padilla, ``Perfect Metamaterial Absorber,'' Physical Review Letters, vol. 100, no. 20, 2008.

\bibitem{Sohl2008Scattering}
C. Sohl, C. Larsson, M. Gustafsson, and G. Kristensson, ``A scattering and absorption identity for metamaterials: Experimental results and comparison with theory,'' Journal of Applied Physics, vol. 103, no. 5, p. 054906, 2008.

\bibitem{Radi2015Thin}
Y. Ra'di C.R. Simovski, and S.A. Tretyakov, ``Thin Perfect Absorbers for Electromagnetic Waves: Theory, Design, and Realizations,'' Physical Review Applied, vol. 3, no. 3, 2015.

\bibitem{Fernandes2019Topological}
D. E. Fernandes and M. G. Silveirinha, ``Topological Origin of Electromagnetic Energy Sinks,'' Physical Review Applied, vol. 12, no. 1, 2019.

\bibitem{Krasnok2019Anomalies}
A. Krasnok, D. Baranov, H. Li, M.-A. Miri, F. Monticone, and A. Al{\`u}, ``Anomalies in light scattering,'' Advances in Optics and Photonics, vol. 11, no. 4, p. 892, 2019.

\bibitem{Gustafsson2020Upper}
M. Gustafsson, K. Schab, L. Jelinek, and M. Capek, ``Upper bounds on absorption and scattering,'' New Journal of Physics, vol. 22, no. 7, p. 073013, 2020.

\bibitem{Schab2020Trade}
K. Schab, A. Rothschild, K. Nguyen, M. Capek, L. Jelinek, and M. Gustafsson, ``Trade-offs in absorption and scattering by nanophotonic structures,'' Optics Express, vol. 28, no. 24, p. 36584, 2020.

\bibitem{qu2022microwave}
S. Qu and P. Sheng, ``Microwave and Acoustic Absorption Metamaterials,'' Physical Review Applied, vol. 17, no. 4, 2022.

\bibitem{Abdelrahman2022How}
M. I. Abdelrahman and F. Monticone, ``How Thin and Efficient Can a Metasurface Reflector Be? Universal Bounds on Reflection for Any Direction and Polarization,'' Advanced Optical Materials, vol. 11, no. 4, p. 2201782, 2022.

\bibitem{Ye2013Ultrawideband}
D. Ye et al., ``Ultrawideband Dispersion Control of a Metamaterial Surface for Perfectly-Matched-Layer-Like Absorption,'' Physical Review Letters, vol. 111, no. 18, 2013.

\bibitem{cao2022backend}
Z. Cao et al., ``Backend-Balanced-Impedance Concept for Reverse Design of Ultra-Wideband Absorber,'' IEEE Transactions on Antennas and Propagation, vol. 70, no. 11, pp. 11217–11222, 2022.

\bibitem{Parameswaran2022ALow}
A. Parameswaran, A. A. Ovhal, D. Kundu, H. S. Sonalikar, J. Singh, and D. Singh, ``A Low-Profile Ultra-Wideband Absorber Using Lumped Resistor-Loaded Cross Dipoles With Resonant Nodes,'' IEEE Transactions on Electromagnetic Compatibility, vol. 64, no. 5, pp. 1758–1766, 2022.

\bibitem{Lin2023Extremely}
Z.-C. Lin, Y. Zhang, L. Li, Y.-T. Zhao, J. Chen, and K.-D. Xu, ``Extremely Wideband Metamaterial Absorber Using Spatial Lossy Transmission Lines and Resistively Loaded High Impedance Surface,'' IEEE Transactions on Microwave Theory and Techniques, pp. 1–10, 2023.

\bibitem{Holloway1997Comparison}
C. L. Holloway, R. R. DeLyser, R. F. German, P. McKenna, and M. Kanda, ``Comparison of electromagnetic absorber used in anechoic and semi-anechoic chambers for emissions and immunity testing of digital devices,'' IEEE Transactions on Electromagnetic Compatibility, vol. 39, no. 1, pp. 33–47, 1997.

\bibitem{Chung2003Modeling}
B.-K. . Chung and H.-T. . Chuah, ``Modeling of RF Absorber for Application in the Design of Anechoic Chamber,'' Progress In Electromagnetics Research, vol. 43, pp. 273–285, 2003.

\bibitem{Mejean2017Electromagnetic}
C. M\'{e}jean et al., ``Electromagnetic absorber composite made of carbon fibers loaded epoxy foam for anechoic chamber application,'' Materials Science and Engineering: B, vol. 220, pp. 59–65, 2017.


\bibitem{Arien2015Study}
Y. Arien, P. Dixon, M. Khorrami, A. Degraeve, and D. Pissoort, ``Study on the reduction of heatsink radiation by combining grounding pins and absorbing materials,'' 2015 IEEE International Symposium on Electromagnetic Compatibility (EMC), 2015.

\bibitem{Xing2022Ultra}
J. Xing et al., ``Ultra-Thin SSPP-Based Sheet for Suppressing Microwave Radiated Emission in SiP Modules,'' IEEE Transactions on Electromagnetic Compatibility, vol. 65, pp. 1–10, 2022.

\bibitem{Khoshniat2021Metamaterial}
A. Khoshniat and R. Abhari, ``Metamaterial Absorbers for Lining System Shield Box and Packaging: Cavity Analysis and Equivalent Material Design,'' IEEE Transactions on Electromagnetic Compatibility, vol. 63, no. 4, pp. 1007–1014, 2021.

\bibitem{Hagglund2012Plasmonic}
C. H\"{a}gglund and S. P. Apell, ``Plasmonic Near-Field Absorbers for Ultrathin Solar Cells,'' The Journal of Physical Chemistry Letters, vol. 3, no. 10, pp. 1275–1285, 2012.

\bibitem{Mulla2015Perfect}
B. Mulla and C. Sabah, ``Perfect metamaterial absorber design for solar cell applications,'' Waves in Random and Complex Media, vol. 25, no. 3, pp. 382–392, 2015.

\bibitem{rozanov2000ultimate}
K. N. Rozanov, ``Ultimate thickness to bandwidth ratio of radar absorbers,'' IEEE Trans. Ant. Prop., vol. 48, no. 8, pp. 1230–1234, 2000.

%
%
%
%
\bibitem{Doane2013Matching}
J.~ P.~ Doane, K.~ Sertel, and J. ~L. ~Volakis, ``Matching bandwidth limits for arrays backed by a conducting ground plane,'' IEEE Trans. Ant. Prop., vol. 61, no. 5, pp. 2511–2518, 2013.

\bibitem{Felsen}
L. B. Felsen and N. Marcuvitz, \emph{Radiation and Scattering of Waves}. Wiely, 1994.

\bibitem{Comm1}
Note that both sides of Eq.~(\ref{Rozanov_Volakis_Bound}) are positive.

%
%
%

\bibitem{Adler1968Electromagnetic}
R. B. Adler, L. J. Chu, and R. M. Fano, \emph{Electromagnetic Energy Transmission and Radiation}, Hoboken, NJ, USA: Wiley, 1968.

\bibitem{Collin1966Foundations}
R. E. Collin, \emph{Foundations for Microwave Engineering}. McGraw-Hill, New York, 1966.

\bibitem{Pozar2011Microwave}
D. M. Pozar, \emph{Microwave Engineering}. 4th ed. John Wiley
$\&$ Sons, Hoboken, NJ, 2011.

\bibitem{Jackson2007Classical}
J. D. Jackson, \emph{Classical Electrodynamics}. John Wiley \& Sons, Hoboken, 2007.

\bibitem{Balanis2016Antenna}
C. A. Balanis, Antenna theory analysis and design. Hoboken, New Jersey Wiley, 2016.

\bibitem{HFSS}
https://www.ansys.com/products/electronics/ansys-hfss

\bibitem{Bybi2008Aquasi}
P.C. Bybi, G. Augustin, B. Jitha , C. K. Aanandan, K. Vasudevan and P. Mohanan, ``A quasi-omnidirectional antenna for modern wireless communication gadgets,'' IEEE Ant. Wire. Propag. Lett, vol.7, 2008.

\bibitem{Shah2021Recent}
S. I. H. Shah, S. M. Radha, P. Park, and I-J Yoon, ``Recent advancements in quasi-isotropic antennas: A review,'' IEEE Access, vol. 9, 2021.

\bibitem{Zhang2022LowProfile}
Z. Zhang, S. Liao, Y. Yang, W. Che and Q. Xue, ``Low-profile and shared aperture dual-polarized omnidirectional antenna by reusing structure of annular quasi-dipole array,'' IEEE Trans. Ant. Prop., vol. 70, no. 9, 2022.

\bibitem{Tretyakov2003Analytical}
Sergei Tretyakov, Analytical modeling in applied electromagnetics. Boston: Artech House, 2003.

\bibitem{firestein2021absorption}
C. Firestein, A. Shlivinski, and Y. Hadad, ``Absorption and Scattering by a Temporally Switched Lossy Layer: Going beyond the Rozanov Bound,'' Phys. Rev. Appl., vol. 17, no. 1, 2022.

\bibitem{Abdul2002Simple}
M. R. Abdul-Gaffoor, H. K. Smith, A. A. Kishk and A. W. Glisson, ``Simple and efficient full-wave modeling of electromagnetic coupling in realistic RF multilayer PCB layouts,''  IEEE Transactions on Microwave Theory and Techniques, vol. 50, no. 6, 2002.

\bibitem{Jones2004PCB}
D. L. Jones, \emph{PCB Design Tutorial}, 2004: \\
chrome-extension://efaidnbmnnnibpcajpcglclefindmkaj/https:\\
//alfadex.com/wp-content/uploads/2014/02/\\
PCBDesignTutorialRevA.pdf


\bibitem{EmbeddedRes}
chrome-extension://efaidnbmnnnibpcajpcglclefindmkaj/https:// \\
koaeurope.de/wp-content/uploads/XR73-Embedded-Resistors-2016-12-15.pdf

\bibitem{Smith2002Determination}
D. R. Smith, S. Schultz, P. Markoš, and C. M. Soukoulis, ``Determination of effective permittivity and permeability of metamaterials from reflection and transmission coefficients,'' Phys. Rev. B, vol. 65, no. 19, 2002.



\bibitem{ROGERS}
https://www.rogerscorp.com/advanced-electronics-solutions


\bibitem{Menzel2008Retrieving}
C. Menzel, C. Rockstuhl, T. Paul, F. Lederer, and T. Pertsch, ``Retrieving effective parameters for metamaterials at oblique incidence,'' Physical Review B, vol. 77, no. 19, 2008.

\bibitem{Cohen2015Bi}
D.~Cohen,R.~Shavit,``Bi-anisotropic metamaterials effective constitutive parameters extraction using oblique incidence S-parameter method,'' IEEE Transactions on Antennas and Propagation, vol. 63, no. 5, 2015.

\bibitem{Watanabe2008S-pol}
R. Watanabe, M. Iwanaga and T. Ishihara, ``s-polarization Brewster's angle of stratified metal–dielectric metamaterial in optical regime,'' Phys. Stat. Sol. (b), vol. 245, 2008.

\bibitem{solt2010Unique}
Z. Szabo, G.-H. Park, R. Hedge, and E.-P. Li, ``A Unique Extraction of Metamaterial Parameters Based on Kramers–Kronig Relationship,'' IEEE Transactions on Microwave Theory and Technique, vol. 58, no. 10, pp. 2646–2653, 2010.





\end{thebibliography}
\end{document}